\documentclass[11pt, oneside]{article} 
\pdfoutput=1

\usepackage{jheppub}

\usepackage{amsmath,amssymb,amsfonts}
\usepackage{color}
\definecolor{darkblue}{rgb}{0.1,0.1,.7}
\usepackage[]{amsmath}
\usepackage[]{graphicx}
\usepackage[]{latexsym}
\usepackage{amscd}
\usepackage[all,cmtip]{xy}
\usepackage{mathrsfs}
\usepackage{bbold}
\usepackage{subfigure}
\usepackage[margin=10pt,font=small,labelfont=bf]{caption}
\usepackage{simplewick}
\usepackage{changepage}
\usepackage[]{algorithm2e}
\usepackage{booktabs,multirow}
\usepackage[OT2,OT1]{fontenc}

\usepackage{slashed}
\usepackage{dsfont}

\def\eps{\epsilon}
\newcommand{\beq}{\begin{equation}} 
\newcommand{\eeq}{\end{equation}}
 
\def\nn{\nonumber}

\def\calN {{\cal N}}

\def\ge{\geqslant}
\def\le{\leqslant}
\def\geq{\geqslant}
\def\leq{\leqslant}
\newcommand{\diffop}[2]{\ifthenelse{\equal{#2}{1}}{\frac{\mrm{d}}{\mrm{d} #1}}{\frac{\mrm{d}^#2}{\mrm{d} #1^#2}}}

\newcommand{\tmtt}[1]{\texttt{#1}}

\def\1{{\ds 1}}

\def\<{\langle}
\def\>{\rangle}

\def\e{\epsilon}
\def\s{\sigma}

\newcommand{\CCBY}{\href{https://creativecommons.org/licenses/by/4.0/}{CC BY 4.0}}

\usepackage{marginnote}
\usepackage[normalem]{ulem}

\newcommand{\be}{\begin{equation}}
	\newcommand{\ee}{\end{equation}}
\newcommand{\bea}{\begin{eqnarray}}
	\newcommand{\eea}{\end{eqnarray}}

\title{New Developments in the Numerical Conformal Bootstrap}
\author{Slava Rychkov$^{a}$, Ning Su$^{b,c,d}$}
\affiliation{$^a$  Institut des Hautes \'Etudes Scientifiques, 91440 Bures-sur-Yvette, France\\
	$^b$ Department of Physics, University of Pisa, I-56127 Pisa, Italy\\
 $^c$ Walter Burke Institute for Theoretical Physics, Caltech, Pasadena, California 91125, USA\\
 $^d$ Department of Physics, Massachusetts Institute of Technology, Cambridge, MA 02139, USA
 }

\abstract{
The numerical conformal bootstrap has become in the last 15 years an indispensable tool for studying strongly coupled CFTs in various dimensions.  Here we review the main developments in the field in the last 5 years, since the appearance of the previous comprehensive review \cite{Poland:2018epd}. We describe developments in the software ({\tt SDPB 2.0}, {\tt scalar\_blocks}, {\tt blocks\_3d}, {\tt autoboot}, {\tt hyperion}, {\tt simpleboot}), and on the algorithmic side (Delaunay triangulation, cutting surface, tiptop, navigator function, skydive). We also describe the main physics applications which were obtained using the new technology.
}

\date{}

\makeatletter
\gdef\@fpheader{}
\makeatother

\begin{document}
\vspace*{-.6in} \thispagestyle{empty}
\begin{flushright}
CALT-TH 2023-046
\end{flushright}

\maketitle

\tableofcontents{}

\section{Introduction}

Conformal field theories (CFTs) are central in modern theoretical physics. They feature prominently as descriptions of continuous thermodynamic \cite{DombGreenVol11} and quantum \cite{Sachdev_2011} phase transitions, and of fixed points of renormalization group (RG) flows \cite{Nakayama:2013is}. They also play an important role in the study of the space of quantum field theories \cite{Douglas:2010ic}, of string theory \cite{Polchinski:1998rq}, of holographic approaches to quantum gravity \cite{Aharony:1999ti}, and of models of particle physics beyond the Standard Model \cite{Luty:2004ye}. A deep feature of CFTs is that their correlation functions can be defined nonperturbatively without explicit reference to a Lagrangian or any other microscopic description. Instead, the main dynamical principle of CFT is the operator product expansion (OPE) \cite{Wilson:1969zs,PhysRevLett.23.1430} which says that the local operators of the theory form an algebra under multiplication. Different CFTs are characterized by the spectrum of the local operators and by the OPE coefficients, which are the structure constants of the operator algebra. Possible OPE algebras are tightly constrained by the constraint of crossing symmetry: the correlation functions of the theory should be unambiguously defined independently of the order in which one may decide to apply the OPE. The program of constraining or solving CFTs using this constraint, called the conformal bootstrap \cite{Polyakov:1974gs},\cite{Belavin:1984vu}, experienced rapid development in the last 15 years.

In this review we focus on the numerical conformal bootstrap, which rewrites the constraint of crossing symmetry as a numerical problem to be solved on a computer \cite{Rattazzi:2008pe}.\footnote{See \cite{Bissi:2022mrs,Hartman:2022zik} for reviews of parallel developments in the analytic conformal bootstrap.} The first 10 years of the numerical conformal bootstrap were thoroughly reviewed in \cite{Poland:2018epd}.

Since then, software and algorithmic advances led to marked increase in the sophistication of performed computations, and to a number of spectacular breakthroughs. Here we review these recent exciting developments. Our review is complementary to another recent review \cite{Poland:2022qrs}. We cover a larger number of topics and go more in-depth in the description of the algorithms and of the results. 

We start by describing the developments in the software allowing to perform the convex optimization, compute conformal blocks, and organize the computations (Section \ref{sec:soft}). We then present a few physics results which were achieved only thanks to the software developments (Section \ref{sec:old}). In the next several section we describe the main algorithmic developments (Delaunay triangulation, surface cutting, {\tt tiptop}, the navigator function, and {\tt skydive}), together with the physics results achieved thanks to them. We then conclude with prospects for future developments.

Our review is directed to an audience which possesses some familiarity with the conformal bootstrap and its vocabulary. We advise others to get acquainted with Sections I-IV of \cite{Poland:2018epd} before proceeding with our review.

\section{Software developments}
\label{sec:soft}

The most important software development was the release of {\tt SDPB} 2.0
{\cite{Landry:2019qug}}, a new version of the main bootstrap code {\tt SDPB}
{\cite{Simmons-Duffin:2015qma}}. {\tt SDPB} implements a primal-dual
interior-point method for solving semidefinite programs (SDPs), taking advantage of
sparsity patterns in matrices that arise in bootstrap problems, and using
arbitrary-precision arithmetic to deal with numerical instabilities when
inverting poorly-conditioned matrices.

The new version has improved the parallelization support. {\tt SDPB} 1.0 could only
run on a single cluster node, using all cores of that node. 
{\tt SDPB} 2.0 can run on multiple cluster nodes with hundreds of cores. The speedup is
proportional to the number of available cores. In addition, every core needs
only a part of the input data, decreasing the amount of needed memory per
node. Provided that one has access to a big cluster, {\tt SDPB} 2.0 can perform
large computations which with {\tt SDPB} 1.0 would take too long to complete, or
would not even fit on a single cluster node due to memory limitations.

\subsection{Conformal blocks}\label{CB}

To setup a bootstrap computation, one needs to precompute $z, \bar{z}$
derivatives of conformal blocks. Up to a prefactor, they can be approximated
by polynomials of the exchanged primary dimension. These polynomials are then
processed by a framework software (Section \ref{framework}) and passed to
{\tt SDPB}.

The simplest blocks are for the scalar
external operators (``scalar blocks''). Many bootstrap studies computed them
used an algorithm from {\cite{El-Showk:2012cjh,El-Showk:2014dwa}}, implemented
in Mathematica. This was limited by the number of Mathematica licenses
available on the cluster. Recently, the algorithm was re-implemented as a fast
open-source C++ program {\tmtt{scalar\_blocks}} {\cite{scalarblocks}}. It
computes scalar blocks in any spacetime dimension $d$, both integer and
non-integer.

Spinning external operators (such as fermions, conserved currents, and the
stress tensor) are playing an increasingly important role in the bootstrap.
Their conformal blocks are more complicated than scalar blocks. Previously,
spinning operator studies relied on Mathematica codes to compute the blocks.
Recently, a C++ program {\tmtt{blocks\_3d}}
{\cite{Erramilli:2019njx,Erramilli:2020rlr}} was released, which computes $d =
3$ conformal blocks for external and exchanged operators of arbitrary spins.

\subsection{Frameworks}\label{framework}

A bootstrap framework is a software suite which, given problem specifications,
generates CFT bootstrap equations, computes $z, \bar{z}$ derivatives of the
crossing equation terms from the conformal block derivatives, generates {\tt SDPB}
input files, launches {\tt SDPB} computations and supervises them. Several frameworks have been developed, the most powerful being {\tt hyperion} \cite{hyperion} in Haskell, and {\tt simpleboot} \cite{simpleboot} in Mathematica, These frameworks also have support for {\tt scalar\_blocks}, {\tt blocks\_3d}, {\tt SDPB} 2.0, as well as the algorithms discussed in later sections of this review (Delaunay triangulation, surface cutting, the navigator function and the skydive). Other frameworks include {\tt juliBootS} \cite{Paulos:2014vya} in Julia, and {\tt PyCFTBoot} \cite{Behan:2016dtz}, {\tt cboot} \cite{cboot} in Python. 

For CFTs with global symmetry, the bootstrap equations involve the crossing kernels (6-j symbols) of the group. For finite groups and Lie groups with a small number of generators, {\tt autoboot} \cite{Go:2019lke,Go:2020ahx} is a package designed to automate the derivation of the crossing kernel and produce bootstrap equations for scalar correlators. The package uses {\tt cboot} as a framework software to launch {\tt SDPB}. The framework {\tt simpleboot} can also work with the bootstrap equations produced by {\tt autoboot}.

\section{Pushing the old method}
\label{sec:old}

In years 2015-2020 numerical bootstrap studies were mostly performed using the following approach \cite{Kos:2014bka,Simmons-Duffin:2015qma}. The bootstrap equations are truncated by taking the derivatives at the crossing symmetric point. Derivatives of the conformal block are approximated by polynomials, up to a positive prefactor. Together with certain assumptions on CFT data, the bootstrap equation is translated into an SDP problem. The SDP is solved in the ``feasibility mode" to determine whether the assumptions are allowed or disallowed. Parameters of the theory space are scanned over using bisection or the Delaunay search (see section \ref{sec:algorithms}). With this method (referred to in this review as ``the old method''), the number of parameters usually cannot be too large. However, many interesting results have been produced in the last few years using this method, especially thanks to the software developments described in section \ref{sec:soft}. In this section, we will review some of those results. Most computations in this section used {\tmtt{SDPB}} 2.0 to solve SDPs, and would not have been possible with {\tmtt{SDPB}} 1.0. 

\subsection{3D $\calN=1$ supersymmetric Ising CFT}

The Gross-Neveu-Yukawa model with one Majorana spinor is described by the following Lagrangian
\be\label{Lag_GNYsuperIsing}
\mathcal{L}=\frac{1}{2}(\partial_{\mu} \sigma)^2+m^2 \sigma^2+\bar{\psi} \slashed{\partial}\psi+\frac{\lambda_1}{2}\sigma\bar{\psi}{\psi}+\frac{\lambda_2^2}{8}\sigma^4,
\ee
where $\psi$ is a three-dimensional Majorana spinor. The Lagrangian is invariant under the time-reversal symmetry (T-parity) where $\sigma\rightarrow -\sigma$ and $\psi \rightarrow \gamma^0\psi$. In the UV, when $\lambda_1=\lambda_2$ and $m=0$, the model becomes $\mathcal{N}=1$ SUSY Wess-Zumino model with superpotential $\mathcal{W}=\lambda \Sigma^3$ and $\Sigma=\sigma+\bar{\theta}\psi+\frac{1}{2}\bar{\theta}\theta \e$. For a particular value of the mass term $m^2$, the model flows in the IR to a fixed point referred to as the 3D $\calN=1$ SUSY Ising CFT. The SUSY is explicitly broken for generic values $\lambda_1 \neq \lambda_2$. However, the SUSY could emerge in the IR from a non-supersymmetric system by tuning a single macroscopic parameter, provided there is only one relevant singlet under T-parity \cite{Grover:2013rc}. Ref. \cite{Grover:2013rc} argued that this fixed point might be realized as a quantum critical point at the boundary of a $3+1$D topological superconductor. 

The 3D $\calN=1$ SUSY Ising CFT was first studied via bootstrap techniques using a single correlator of a scalar \cite{Bashkirov:2013vya} or a fermion \cite{Iliesiu:2015qra} (see \cite{Poland:2018epd} for a review). Later, Refs. \cite{Rong:2018okz,Atanasov:2018kqw} extended the analysis to correlators involving $\sigma,\epsilon$. From a non-SUSY point of view, the bootstrap setup is same as the 3D Ising $\sigma,\epsilon$ setup of \cite{Kos:2014bka}, where $\mathbb{Z}_2$ is realized by T-parity. Supersymmetry imposes additional constraints: $\Delta_{\epsilon}=\Delta_{\s}+1$, since $\s$ and $\epsilon$ are in the same supermultiplet, and the OPE coefficients $\lambda_{\s\s  O}$, $\lambda_{\e\e  O}$ are related to $\lambda_{\s\e  P}$, where $O$ and $P$ belong to the same supermultiplet. Ref. \cite{Rong:2018okz} thoroughly worked out those constraints and injected them into the $\sigma,\epsilon$ setup. By demanding that, under T-parity, only one even scalar operator and two odd scalar operators are relevant, Ref. \cite{Rong:2018okz} isolated the CFT as a small island in the theory space and produced precise critical exponents with rigorous error bars: $\eta_{\s}=0.168888({\bf 60})$ and $\omega=0.882({\bf 9})$.\footnote{Prior to this work, the best estimates of those critical exponents were $\eta_{\s}=0.170$ and $\omega=0.838$, obtained from the four loop $\epsilon$-expansion of the Gross-Neveu-Yukawa model \cite{Zerf:2017zqi}. Due to the sign problem, there is no Monte Carlo simulation for this model.} This provides strong evidence that the theory has superconformal symmetry and has only one relevant singlet under T-parity, confirming the possibility of emergent supersymmetry.

\begin{figure}[t!]
\centering
\includegraphics[width=0.48\textwidth]{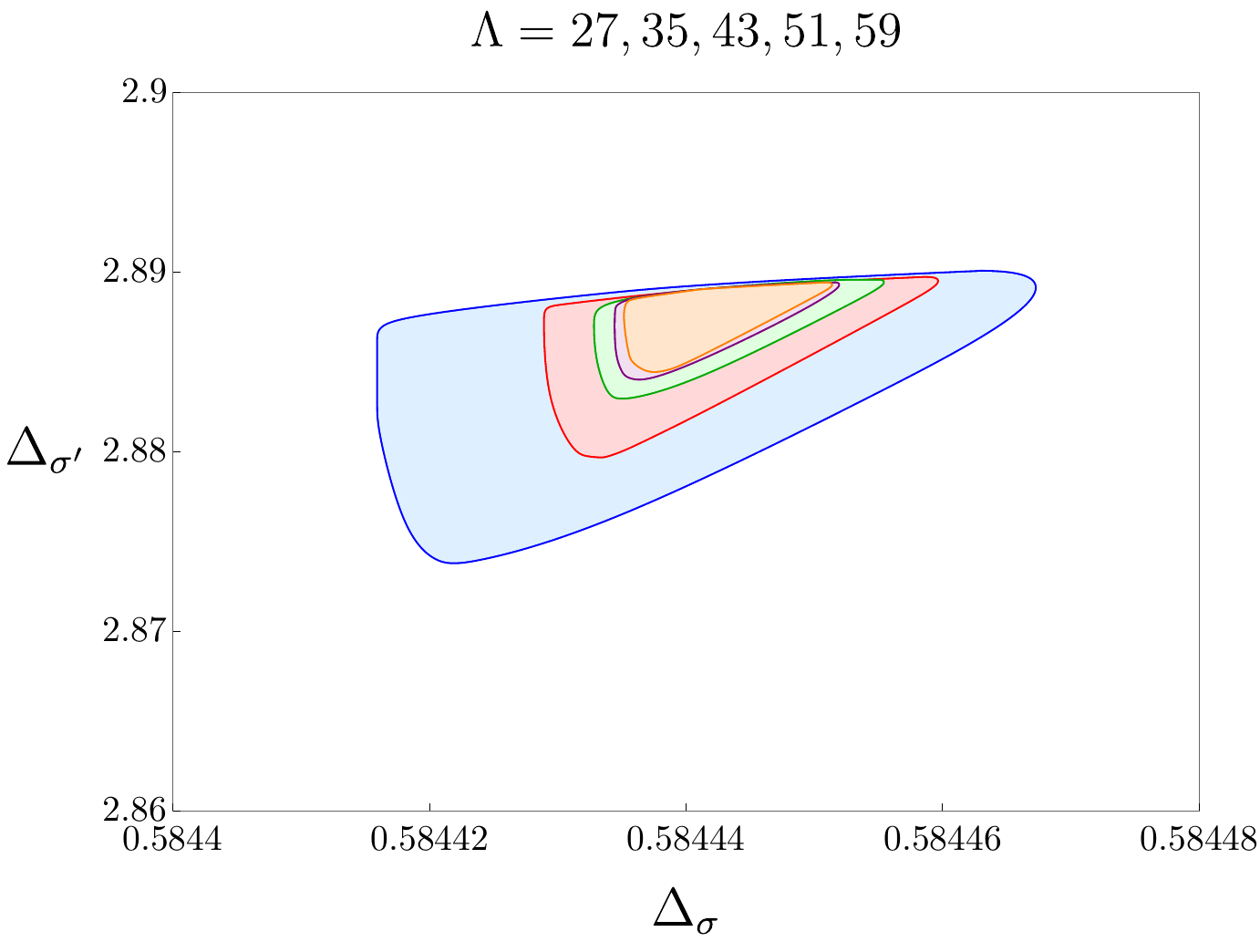}
\caption{\label{fig:superIsing}
(Color online) Allowed region from \cite{Atanasov:2022bpi} for the scaling dimensions of the leading parity-odd scalars ${\sigma, \sigma'}$
in the 3d $\calN=1$ super-Ising model. \CCBY.}
\end{figure}

There are two follow up works \cite{Rong:2019qer,Atanasov:2022bpi}. As of 2023, the most precise conformal data for this CFT was obtained from \cite{Atanasov:2022bpi}, using the same setup as \cite{Rong:2018okz} but at a much higher $\Lambda=59$. The computation at such a high $\Lambda$ becomes practically feasible by utilizing \tmtt{scalar\_blocks} and \tmtt{SDPB} 2.0. The result is summarized in Table \ref{tab:superIsingresult} and Figure \ref{fig:superIsing}.\footnote{The relations between scaling dimensions and the critical exponents are as follows: 
$\eta_{\sigma} = 2 \Delta_\sigma +1$; $1/\nu = 1- \Delta_\epsilon$, where $\Delta_\epsilon=\Delta_\sigma+1$; $\omega=\Delta_{\epsilon'}-3$, where $\epsilon'$ is the leading irrelevant singlet. 
In the present case, $\epsilon'$ is the bosonic descendant of $\sigma'$, and $\Delta_{\epsilon'}=\Delta_{\sigma'}+1$.}

\begin{table}[htp]
	\centering
	\begin{tabular}{@{}|c|c|@{}}
		\hline
		CFT data  & critical exponents\\
		\hline
		$\Delta_\sigma$=0.5844435({\bf 83}) & 
		\(\begin{aligned}
			\eta_{\sigma} = \eta_{\psi} &= 0.168887({\bf 17})\\
			1/\nu &= 1.415557({\bf 8})
		\end{aligned}\) \\ 
		$\Delta_{\sigma'}$=2.8869({\bf 25}) & $\omega = 0.8869({\bf 25})$  \\
		\hline
	\end{tabular}
	\caption{Results of \cite{Atanasov:2022bpi} for the scaling dimensions of the leading parity-odd scalars ${\sigma, \sigma'}$ in in the 3d $\calN=1$ super-Ising model. Uncertainties in bold are rigorous.
		\label{tab:superIsingresult}}
\end{table}

The work \cite{Rong:2019qer} studied more general $\mathcal{N}=1$ Wess-Zumino models with global symmetries, obtaining strong bounds on CFT data. However, no small bootstrap islands have been identified. The bootstrapping of generic SUSY Wess-Zumino models remains a challenging task.

\subsection{Bootstrapping critical gauge theories}

An important class of CFTs is critical gauge theories, i.e.~IR fixed point of abelian or non-abelian gauge fields coupled to matter fields. The simplest examples are fermionic and bosonic 3D Quantum Electrodynamics ($\text{QED}_3$), where $N_f$ Dirac fermions or complex bosons are coupled to the $U(1)$ gauge field. In both cases, it is believed that the theories will flow to an IR CFT at large $N_f$, while in the small $N_f$ cases, they might not be critical. Determining the precise extent of the conformal window of these theories (i.e.~the range of $N_f$ when they flow to an IR fixed point) presents a longstanding problem.\footnote{See \cite{Gukov:2016tnp}, Section 4, for a survey of approaches to the conformal window problem for the fermionic $\text{QED}_3$.} Moreover, these models have a rich connection to the Deconfined Quantum Critical Point (DQCP) \cite{deccp,deccplong} and Dirac Spin Liquid (DSL) \cite{Hermele_2005,Song:2018ial,Song:2018ccm} in condensed matter physics.

Critical gauge theories present interesting targets for the conformal bootstrap. Previously, monopole operators in fermionic $\text{QED}_3$ were studied via the bootstrap in \cite{Chester:2016wrc,Chester:2017vdh}. Various bootstrap bounds for bosonic $\text{QED}_3$ were obtained in \cite{Nakayama:2016jhq,Iliesiu:2018,Li:2021emd}, offering insights into the nature of DQCP (see also \cite{Poland:2018epd,Poland:2022qrs} for a review). In this section, we will discuss more recent progress, focusing on 3D non-supersymmetric theories.\footnote{We refer the reader to \cite{Chester:2016wrc,Chester:2017vdh} for progress on 4D critical gauge theories}

Critical gauge theories are, in general, more difficult to bootstrap than the fixed points of scalar theories, like the $O(N)$ model. One difficulty arises because the fundamental matter field is charged under the gauge group, and hence their correlation functions are not good CFT observables. Gauge-invariant operators are built out of products of fundamental fields and have a higher scaling dimension, while the numerical bootstrap is known to converge slower when operators of higher scaling dimensions are involved. Because of this and other difficulties, small bootstrap islands have not yet been obtained even for the simplest critical gauge theories. New ideas may be needed, such as bootstrapping correlators of closed or open Wilson lines. For the moment it is not clear how to effectively bootstrap these objects. Below we will discuss works which bootstrapped correlators of gauge-invariant operators, using the ``old method''. 

The global symmetries of $\text{QED}_3$ are the $SU(N_f)$ flavor symmetry of the matter sectors, as well as the topological $U(1)$ symmetry. The monopole operators carry a non-zero charge $q$ under $U(1)$, and their specific representation under $SU(N_f)$ depends on $q$, $N_f$, and whether the matter sector is bosonic or fermionic. Therefore, the bootstrap equations involving monopoles are different for bosonic and fermionic $\text{QED}_3$. In Sections \ref{sec:fQED} and \ref{sec:bQED}, we discuss the bootstrap studies of scalar correlators in fermionic and bosonic $\text{QED}_3$, respectively. In Section \ref{sec:JJJJ}, we discuss bootstrap studies of the correlator of flavor currents. This setup applies to all CFTs with a flavor current.

\subsubsection{Bootstrapping fermionic $\text{QED}_3$} \label{sec:fQED}

It is believed that fermionic $\text{QED}_3$ at large $N_f$ flows to a CFT with a global symmetry $G_{\rm IR}=SU(N_f) \times U(1)$ in the IR, where $SU(N_f)$ is the flavor symmetry of the four Dirac fermions and the $U(1)$ is the flux conservation symmetry of the gauge field. At small $N_f<N_\text{crit}$, the IR is in a chiral symmetry breaking phase. The specific value of $N_\text{crit}$ has been studied using many different methods, yet there is no consensus \cite{Gukov:2016tnp}.

The $N_f=4$ case is particularly interesting. If the $N_f=4$ case is indeed conformal, several lattice models and materials, including spin-1/2 Heisenberg model on the Kagome and triangular lattices, are conjectured to realize it as a critical conformal phase \cite{He2017, Hu2019}. The UV lattice models usually possess a symmetry $G_{\rm UV}$ which is smaller than $G_{\rm IR}$. For the fixed point to be reached, i.e.~for the conformal phase to be realized in those lattice models, all $G_{\rm UV}$ singlet operators (which includes $G_{\rm IR}$ singlets but also some operators charged under $G_{\rm IR}$) have to be irrelevant. Therefore the fate of those lattice models depends on the precise spectrum data of the IR CFT.

\begin{figure}[t!]
	\centering
	\includegraphics[width=0.45\textwidth]{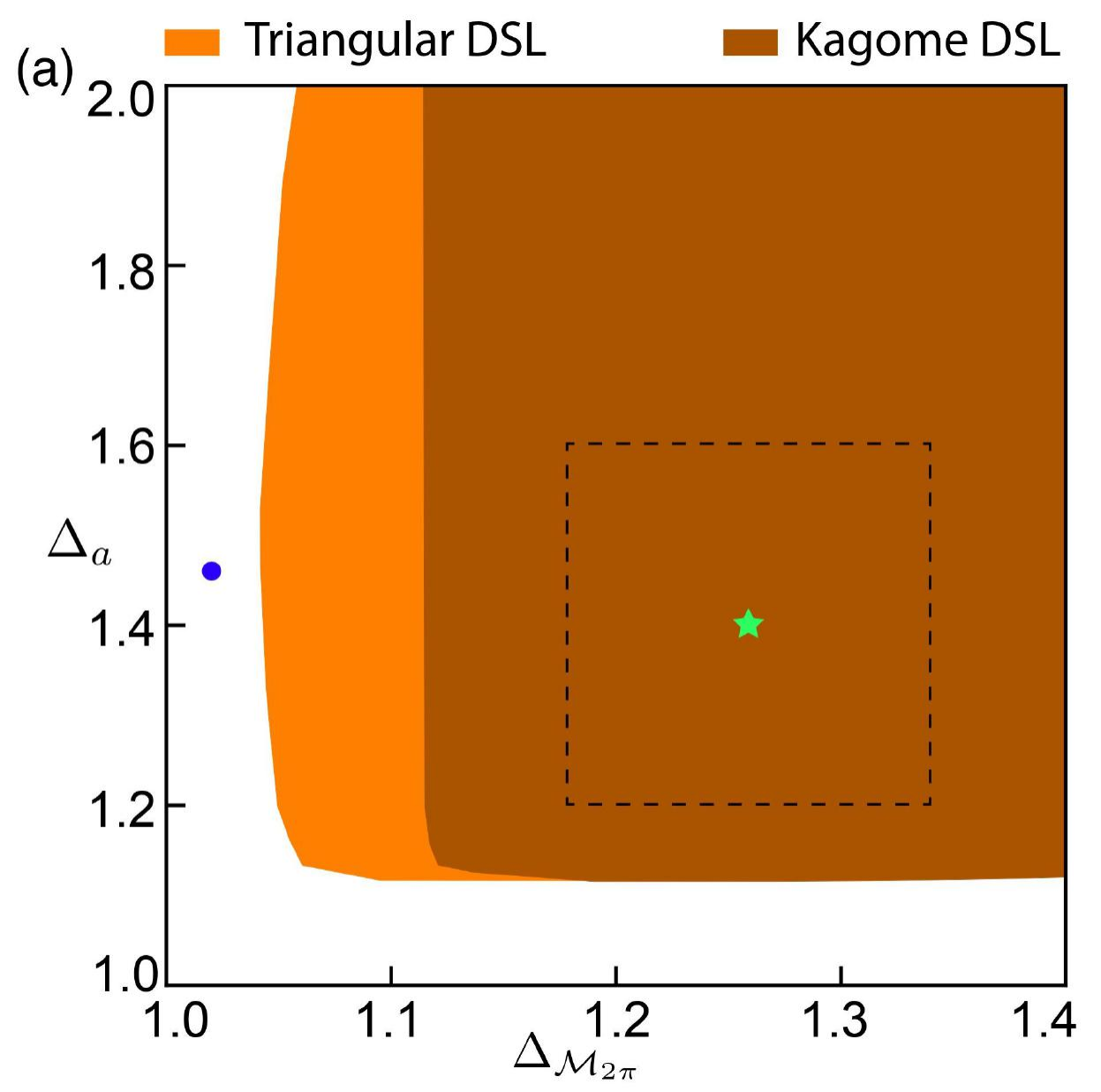}
	\includegraphics[width=0.45\textwidth]{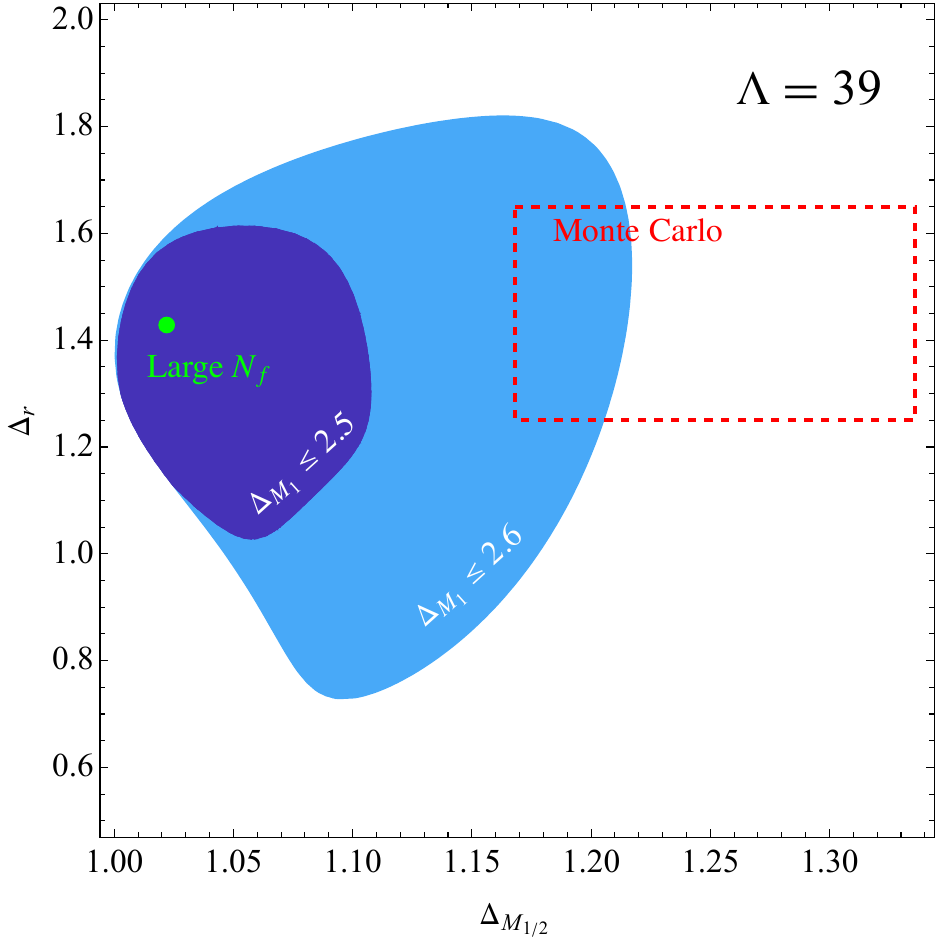}
	\caption{
		(Color online) On the top: The allowed region in the space of dimensions of the fermion bilinear ($a$) and the monopole ($\mathcal M_{2\pi}$) from Ref.~\cite{He:2021sto}, \CCBY.  The gap assumptions are imposed to be compatible with the conformal phase scenario on triangular lattice and the Kagome lattice. The green (light gray) star and the dashed error box represent the results of a Monte Carlo simulation, while the blue point is from the large-$N_f$ expansion. On the bottom: The allowed region in the space of dimensions of the fermion bilinear ($\Delta_r$) and the monopole ($\mathcal M_{1}$) from Ref. \cite{Albayrak:2021xtd}. Here, the green (light gray) dot indicates the result from the large-$N_f$ expansion, while the red dashed box is from a Monte Carlo simulation. See Figure 5 of \cite{Albayrak:2021xtd} for details. 
	}
	\label{fig:Nf4QED3}
\end{figure}

Specifically, the topological charge $q=1$ monopole operator $M_1$ and a charge neutral operator\footnote{Here $(abc)$ denotes the Young diagram of the representation. Both $M_1$ and $S_{(220)}$ transforms as $(220)$ under $SU(4)$. In Ref. \cite{He:2021sto}, $M_1$ is referred to as $M_{4\pi}$ or $(T,T)$, and $S_{220}$ is called $(T,S)$, where $T$ and $S$ refer to the traceless symmetric tensor and singlet in $O(6)\times O(2) \approx SU(N_f) \times U(1)$.} $S_{(220)}$ 
must be irrelevant for the conformal phase scenario to be realized in the Kagome lattice \cite{Song2018,Song2018a}. $\text{QED}_3$ has been studied using both large-$N_f$ expansion and Monte Carlo simulations. While Monte Carlo simulation support the CFT scenario \cite{Karthik2015, Karthik2016,Karthik2019}, large-$N$ prediction yield $\Delta_{M_1}\approx 2.499$ and $\Delta_{S_{(220)}}\approx 2.379$ \cite{Chester:2016wrc}, which are relevant and would rule out the CFT scenario on the Kagome Lattice.

The bootstrap study of a single correlator of the fermion bilinear operator was initiated in \cite{Li:2018lyb}, where various large-$N$ expansion results were injected into the setup. That study observed a kink moving with the dimension $\Delta_*$ of the first scalar in a certain representation (denoted as $T\bar{A}$ in \cite{Li:2018lyb}), 
and suggested that the kink may correspond to the fermionic $\text{QED}_3$ if $\Delta_*$ is fixed to the actual CFT value.

After that, two bootstrap studies \cite{He:2021sto, Albayrak:2021xtd} investigated the mixed correlators between the lowest monopole and the fermion bilinears. Strong bounds on CFT data have been obtained; see Fig. \ref{fig:Nf4QED3}. Ref. \cite{He:2021sto} imposed gap assumptions that are compatible with the conformal phase scenarios (specifically $\Delta_{M_1} \ge 3$ and $\Delta_{S_{(220)}}\ge 3$ for the Kagome lattice). The resulting bounds are consistent with the latest Monte Carlo result but excluded certain results from large $N_f$ expansion. On the other hand, Ref. \cite{Albayrak:2021xtd} imposed various gap assumptions inspired by the large $N_f$ result. Specifically it assumed $\Delta_{M_1}\le 2.6, \Delta_{S_{(220)}}\ge 2.8$. They found bounds in an isolated region compatible with the large-$N$ prediction. This work also introduced a useful technique called interval positivity to implement gap assumptions of the form $\Delta_\text{min} \le \Delta \le \Delta_\text{max}$. In both works, the (ir)relevance of $M_1$ and $S_{(220)}$ is not determined by bootstrap, but inputted as an assumption, and a reliable spectrum has not been obtained. Thus the matter cannot be considered settled. 

For $N_f=2$ case, the work \cite{Li:2021emd} bootstrapped the correlator of the lowest monopole operator. Imposing the constraints of RG stability and/or of the $O(4)$ symmetry enhancement \cite{2015PhRvB..92v0416X}, the bounds of \cite{Li:2021emd} were confronted with determinations of scaling dimensions in Monte Carlo simulations, being inconsistent with \cite{Qin:2017cqw}, while marginally consistent with \cite{Karthik:2016ppr,Karthik:2019mrr}.

\subsubsection{Bootstrapping bosonic $\text{QED}_3$}
\label{sec:bQED}

Several works have bootstrapped the bosonic $\text{QED}_3$. One issue is how to distinguish $\text{QED}_3$ from non-abelian $SU(N_c)$ gauge theories with the same number of matter field multiplets. This question was investigated in \cite{Reehorst:2020phk,He:2021xvg}. The key observation is that there are natural gaps in the spectrum that distinguish between different $N_c$ values. The simplest example concerns an operator of the form $\bar{\phi}^{[f_1}_{[c_1} \bar{\phi}^{f_2]}_{c_2]} {\phi}^{[c_1}_{[f_3} {\phi}^{c_2]}_{f_4]}$, where $\phi$ are the bosonic matter fields in the fundamental of $SU(N_c)$, and $f_i$, $c_i$ are flavor and color indices respectively. This operator exists when $N_c \ge 2$, but an operator transforming in the same way under the global symmetry cannot be constructed in the abelian cases, using just four scalars without derivatives. Thus, imposing a gap in this symmetry sector, in principle, could isolate the abelian case from $N_c \ge 2$. With such a gap assumption, Ref.~\cite{He:2021xvg} bootstrapped the correlator of the leading scalar in $SU(N_f)$ adjoint and found bounds in closed region for the large $N_f$ cases and in (2+$\epsilon$)D for small $N_f$. However, it's not clear whether the closed region contains only the target theory, not anything else, and whether it could converge to a small bootstrap island at large $\Lambda$ and produce precise scalar QED spectrum. 

The $N_f=2$ case is believed to be the simplest example of the Deconfined Quantum Critical Point (DQCP). The Monte Carlo simulations are in tension with the bootstrap bounds (see Ref.~\cite{Poland:2018epd} for a review). A possible way to reconcile the bootstrap results with the Monte Carlo is proposed in Ref.~\cite{Nahum:2019fjw,Ma:2019ysf}. This work suggested a formal interpolation between the 2D $SU(2)_{k=1}$ WZW theory and the 3D DQCP. With a one-loop calculation, the work found the critical dimension $d_c \approx 2.77$, where the theory annihilates with another fixed point and becomes a (pair of) complex CFTs, an example of the merger and annihilation scenario \cite{gorbenko2018walking,Gorbenko:2018dtm}. If this scenario is correct, the bosonic QED$_3$ for $N_f=2$ does not exist in 3D as a unitary fixed point, and the DQCP phase transition is first order but only weakly so because the RG flow passes a long time flowing between a pair of complex CFTs, behavior referred to as ``walking''.

This scenario may be checked via the numerical bootstrap analysis of the $SU(2)_{k=1}$ WZW theory in 2D and in non-integer $d>2$. The first 2D analysis was performed in \cite{Ohtsuki:thesis}, which discovered that the WZW theory sits at a sharp kink of a single correlator bound. Then, Refs.~\cite{Li:2020bnb,He:2020azu} investigated the fate of this kink in non-integer $d>2$, as well as in the case of larger $N_f$. Ref.~\cite{He:2020azu} found evidence to support the critical dimension proposed in Ref.~\cite{Ma:2019ysf}.

In spite of these developments, the puzzle of DQCP cannot be considered fully clarified. Recently, evidence has emerged from Monte Carlo \cite{Takahashi:2024xxd} and from the calculations on the fuzzy sphere\footnote{A novel numerical method proposed in \cite{Zhu:2022gjc}.} \cite{2023arXiv230705307C},\cite{Chen:2024jxe} that the phase transition is weakly first order due to a tricritical point---a scenario more mundane than complex CFT (walking), but previously deemed unlikely as requiring tunings in the microscopic models. (At the same time, another fuzzy sphere calculation \cite{Zhou:2023qfi} argued for consistency with the walking scenario.) The tricritical scenario was considered in a bootstrap study \cite{Chester:2023njo} using more modern techniques, see Section~\ref{sec:skydive-alg}.

\subsubsection{Bootstrapping the flavor currents} \label{sec:JJJJ}

\begin{figure}[t!]
	\centering
	\includegraphics[width=0.45\textwidth]{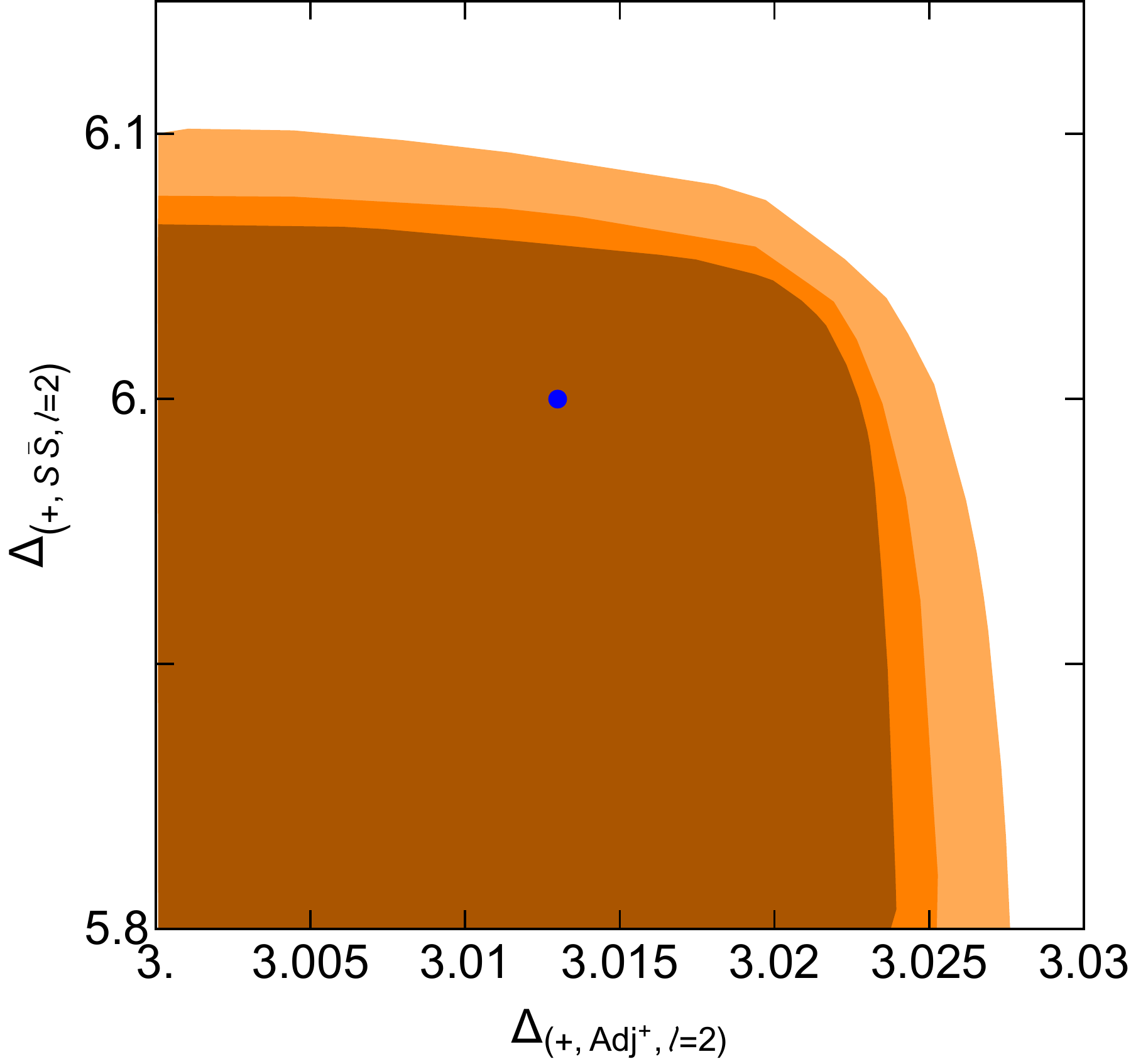}
	\includegraphics[width=0.45\textwidth]{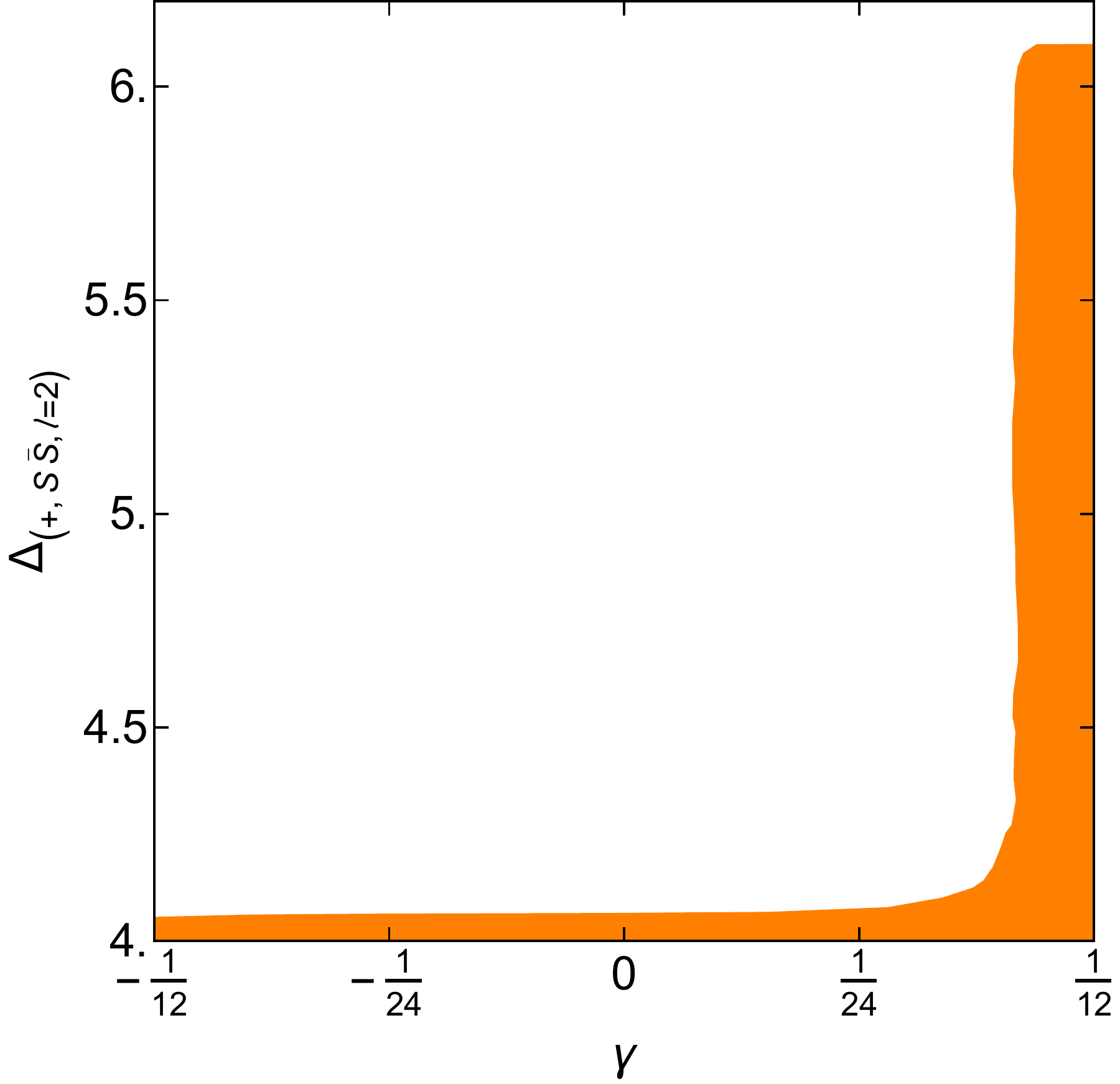}\label{fig:SSbl2_vs_gamma}
	\caption{(Color online) Feasibility bounds from bootstrapping non-abelian currents, Ref.~\cite{He:2023ewx}. On the top: The lowest operator in the $(+, S\bar S$, $\ell=2)$ sector v.s. the sector $(+, Adj^+, \ell=2)$ in $SU(100)$ CFT, where $Adj^+$ is the adjoint representation. The light, medium, and dark orange bounds are for $\Lambda=19,23,27$, respectively. Blue (dark gray) dot is the large-$N$ result. On the bottom: $\gamma$ versus the lowest operator in the $(+, S\bar S$, $\ell=2)$ in $SU(100)$ CFT. $\Lambda=19$. $\gamma$ is a parameter appearing in the current, current, stress tensor 3-point function, and it obeys the conformal collider bounds $|\gamma|\leq 1/2$ \cite{HofmanConformalCollider2008}.  In both cases, mild gaps in several sectors are imposed. The bounds are insensitive to those gap assumptions.}
	\label{fig:JJJJ_SSbl2}
\end{figure}

The existence of conserved currents is a key feature of all local CFTs possessing continuous global symmetry. Global symmetry currents (hereafter referred as simply currents) are CFT primaries of spin 1 which are conserved and have the protected scaling dimension $d-1$. In addition, OPE coefficients of currents with other primaries are constrained by Ward identities. One expects that implementing those properties in a bootstrap setup could lead to new bounds and new spectral data that cannot be accessed in setups dealing only with scalar primaries. 

For the problem of bootstrapping critical gauge theories, correlators involving currents are appealing to consider for several other reasons. For the abelian gauge theories in 3D, the existence of topological current $j_{\text{top}}^{\rho} = \epsilon^{\mu\nu\rho} \partial_{\mu} A_{\nu}$ is a key feature; operators charged under this current are called monopole operators. The existence of this current could e.g.~distinguish fermionic QED$_3$ from the free fermion theory. Below we will see other examples of how correlators of currents can help distinguish critical gauge theories from other theories.

The bootstrap study of the four-point function of currents in 3D CFTs with a $U(1)$ symmetry was initiated in \cite{Dymarsky:2017xzb}. Subsequently, the work \cite{Reehorst:2019pzi} studied the 3D $O(2)$ model using mixed correlators of the $O(2)$ current and a singlet. Ref.~\cite{Reehorst:2019pzi} obtained several new data that was not accessible in prior scalar correlator studies of the $O(2)$ model \cite{Kos:2013tga} (although \cite{Reehorst:2019pzi} used some results of \cite{Kos:2013tga} as an input). As already mentioned, abelian 3D gauge theories have $U(1)$ topological currents. However, no features associated with these theories were observed in the bounds of \cite{Dymarsky:2017xzb,Reehorst:2019pzi}.

More recently, Ref.~\cite{He:2023ewx} bootstrapped for the first time the four-point functions of currents in the CFTs with a non-abelian continuous symmetry. Similar to the bosonic QED case studied in \cite{He:2021xvg}, Ref.~\cite{He:2023ewx} proposed to impose gap assumptions on the spectrum of exchanged operators, which can be used to distinguish between different $N_c$ for the fermionic case, and in particular the abelian from the non-abelian cases. In fermionic $U(N_c)$ gauge theories, an operator of the form $\bar{\psi}^{(f_1, c_1} \gamma^{\mu} \psi_{(f_2, c_1} \bar{\psi}^{f_3, c_2)} \gamma^{\nu} \psi_{f_4, c_2)}$, where $f_i$, $c_i$ denote the flavor and color indices, respectively, and the parentheses symmetrize the $f_i$ flavor indices. This operator exists only for $N_c \ge 2 $ and has a scaling dimension of 4 in the large-$N_f$ limit. Such an operator has scaling dimension 4 in the large-$N_f$ limit. Thus, a gap beyond 4 in this channel could, in principle, distinguish fermionic QED from fermionic QCD. The representation to which this operator belongs is denoted in \cite{He:2023ewx} as $(+, S\bar S$, $\ell=2)$, where $+$ denotes spacetime parity and $\ell$ is the spin. This gap could also remove the solution of Generalized Free Field (GFF), where the leading operator $(J^{\mu })^{(f_1}{}_{(f_2}(J^{\nu})^{f_3)}{}_{f_4)}$ has scaling dimension 4.  Note that $(+, S\bar S$, $\ell=2)$ could be accessed already in the scalar bootstrap of Ref.~\cite{He:2021sto, Albayrak:2021xtd}. However, in that setup, the GFF operator has a scaling dimension of 6.\footnote{To construct a spin 2 operator using scalar operators, two derivatives have to be inserted.} Indeed, Ref.~\cite{He:2021sto} found no signal of $N_f=4$ $\text{QED}_3$ in this sector. We see a clear advantage of the setup involving external currents..

By scanning over the gaps in $(+, S\bar S$, $\ell=2)$ and other parameters, Ref.~\cite{He:2023ewx} obtained strong bounds on CFT data, and observed kinks; see Fig.~\ref{fig:JJJJ_SSbl2}. Large $N_f$ $\text{QED}_3$ appears to be close to some kinks. Ref.~\cite{He:2023ewx} also imposed stronger conditions by demanding that many operators lies within mild windows around the large-$N_f$ values. With such assumptions, in large $N_f$ cases, the bound turns into an isolated region, which does not contain any obvious known theories other than $N_f=4$ $\text{QED}_3$. However, the isolated region is not small enough to produce a precise QED spectrum. Furthermore, it's not clear whether the closed regions could converge to a small bootstrap island at high $\Lambda$.

The discussed recent computations were allowed by the progress in the conformal block and SDP software. In the first $U(1)$ current study of \cite{Dymarsky:2017xzb}, conformal block of the current correlator were computed by decomposing them into (derivatives of) scalar blocks. On the other hand, the setup of \cite{He:2023ewx} produced SDPs that are much larger in size than those in \cite{Dymarsky:2017xzb} and are more challenging to generate and compute. These calculations have become feasible due to the development of \texttt{blocks\_3d} and \texttt{SDPB} 2.0, as discussed in Section \ref{sec:soft}.

To summarize, bootstrapping critical gauge theories remains a challenge. Various attempts to bootstrapping QED in 3D and (2+$\epsilon$)D have been made. Despite having found strong bounds in the theory space, no one has obtained a small bootstrap island that could produce precise and reliable CFT data. In future work,  for a successful bootstrap analysis of QED, it might be necessary to consider larger systems of mixed correlators. One could also consider correlators involving open or closed Wilson lines, although formulating a conformal bootstrap program for these observables is a wide open problem.

\subsection{Multiscalar CFTs}
\label{sec:scalar}

In this review, by a multiscalar CFT we mean an IR fixed point of a Lagrangian field theory of $N$ scalar fields with quartic interactions in the UV, which respect global symmetry $G\subset O(N)$. Physically, one is most interested in stable fixed points, i.e. possessing only one relevant singlet scalar (the mass term). For small $N$, the possible stable fixed points are limited. See \cite{Rong:2023xhz} for a classification of all stable fixed points for up to five scalars, based on one-loop calculations in $d=4-\epsilon$ dimensions. Apart from the Ising and $O(N)$ CFTs, simple multiscalar CFTs include those with symmetries $\mathds Z_2^n \rtimes {\mathcal S_n}$ (the cubic symmetry), $\mathds Z_2 \times {\mathcal S_n}$, $O(m)^n \rtimes {\mathcal S_n}$, and $O(m)\times O(n)$. For all these symmetry groups there exist a family of $d$-dimensional CFTs which for $d=4-\epsilon$ and $\epsilon\ll 1$ are weakly coupled and smoothly connect to the free theory for $d=4$. These families can be studied via the $\epsilon$-expansion which, when extended for $\epsilon=1$ and Borel-resummed, gives predictions for 3D CFTs. Several works have studied CFTs with these global symmetries using the conformal bootstrap. One hope of these studies is to recover the results predicted via the $\epsilon$-expansion using the non-perturbative technique (and perhaps with a better precision). In addition, one may hope to discover other CFTs having the same global symmetry (in addition to the one predicted by the $\epsilon$-expansion).

Bootstrap studies of CFTs with the cubic symmetry $\mathds Z_2^n \rtimes \mathcal S_n$ began with \cite{Rong:2017cow,Stergiou:2018gjj}. Bounds from a single correlator of an operator $v$ in the standard $n$-dimensional representation have been obtained, working in 3D. While some kinks were observed on the bounds, they do not correspond to ``the cubic CFT'', i.e. the CFT predicted by the $\epsilon$-expansion \cite{aharony1973critical}.
 Ref.~\cite{Stergiou:2018gjj} conjectured that the kink corresponds to a new CFT which they dubbed ``Platonic CFT.'' \footnote{ Ref.~\cite{Stergiou:2018gjj} provided evidence that the Platonic CFT exists, and is distinct from the cubic CFT, not only in $d=3$ but also in $d=3.8$. However, from recent work \cite{Rong:2023xhz}, near $d=4$, and with up to five scalars, we don't expect other weakly-coupled theories exhibiting cubic symmetry, apart from the cubic CFT. This suggests that the Platonic CFT might have a larger symmetry and its Lagrangian description may involve more than five scalar fields. It would be interesting to clarify this issue.} Refs.~\cite{Kousvos:2018rhl,Kousvos:2019hgc} further investigated the conjectured Platonic CFT by examining four-point functions involving $v$ and one additional operator. With certain gap assumptions, they constrained the theory to an isolated region.

The works of \cite{Nakayama:2014lva,Henriksson:2020fqi} studied theories with $O(m)\times O(n)$ symmetry. Kinks were observed on the bounds from a single correlator of $\phi_{ar}$ (the bifundamental representation, with $a,r$ being $O(m), O(n)$ vector indices), and they are consistent with the large $n$ expansion for $O(m)\times O(n)$ Wilson-Fisher theories at fixed $m$. Using a mixed correlator setup involving $\phi_{ar}$ and the leading singlet operator, \cite{Henriksson:2020fqi} showed that the CFT can be constrained to an isolated region by demanding certain operator saturate a bound. 

Similar methods were applied to the $O(m)^n \rtimes \mathcal S_n$ symmetry \cite{Stergiou:2019dcv,Henriksson:2021lwn,Kousvos:2021rar} and the $U(m)\times U(n)$ symmetry \cite{Kousvos:2022ewl}. Ref. \cite{Stergiou:2019dcv} studied a single correlator of $\phi_i$, where $i=1,...,mn$ labels $n$ copies of $O(m)$ vectors transforming under $O(m)^n \rtimes \mathcal S_n$. Focusing on the $O(2)^2 \rtimes \mathcal S_2$ case, Ref. \cite{Stergiou:2019dcv} made interesting observations: Several materials are supposed to undergo phase transitions described by a CFT with $O(2)^2 \rtimes \mathcal S_2$. Experiments on these phase transitions have yielded two sets of critical exponents. Surprisingly, two kinks were exhibited on the single correlator bound, which are in good agreement with the two sets of experimental data. Ref. \cite{Henriksson:2021lwn} further studied the kinks at the large-$m$ limit. Ref. \cite{Kousvos:2021rar} investigated a mixed correlator system involving $\phi$ and an operator $X$ that is a singlet under $O(m)$ but a fundamental representation in $\mathcal S_n$. Ref. \cite{Kousvos:2021rar} found isolated regions under certain gap assumptions. However, the fate of the two possible $O(m)^n \rtimes \mathcal S_n$ CFTs remains inconclusive. 

In conclusion, despite obtaining many strong bounds for various multiscalar CFTs, small bootstrap islands have still not yet been obtained for those theories. In an upcoming work \cite{ono2talk,ono2}, this is achieved for the multiscalar CFTs with $O(N)\times O(2)$ global symmetry.

Outside the scope of local theories, \cite{Behan:2018hfx} studied all three relevant scalars in the long-range Ising model and found a kink in the numerical bounds that corresponds to the target CFT; further interesting results on this model were obtained in \cite{Behan:2023ile}.

\subsection{Extraordinary phase transition in the $O(N)$ boundary CFT}

Conformal defects are difficult to bootstrap using the rigorous numerical conformal bootstrap methods based on SDPs, because the OPE coefficients in the bootstrap equation involving both the bulk and the defects generally do not exhibit positivity properties. One non-rigorous method usually applied to bootstrap these systems is Gliozzi's method \cite{Gliozzi:2013ysa}, where one solves directly the truncated bootstrap equation, without assuming positivity of the OPE coefficients. However, the error in this method is not systematically controllable.

In this section, we review a recent exploration \cite{Padayasi:2021sik} on bootstrapping the boundary of 3D $O(N)$ CFTs, representing a special case in which the standard SDP approach can be applied. The system under consideration is the Heisenberg model on a $d$-dimensional lattice with an infinite plane boundary at $x_d=0$:
\begin{align}
	\label{eq:HboundaryCFTON}
	{\cal H } = -\sum_{\langle i,j\rangle } K_{i,j} \vec{S}_i \cdot \vec{S}_j,
\end{align}
where $\vec{S}_i$ is the classical $O(N)$ spin, $\langle i,j\rangle$ denotes a pair of neighboring sites, and $K_{i,j}=K_1$ when the pair is on the boundary and $K_{i,j}=K$ when the pair is in the bulk. Depending on $N$, the system exhibits different properties. An extraordinary-log\footnote{``Extraordinary'' here refers to the enhanced boundary interactions needed to realize this transition, as opposed to the ``ordinary'' transition happening when boundary interactions have the same strength as the bulk ones. This terminology is the same as for the Ising case $N=1$. For $N\ge 2$ the transition is referred to as ``extraordinary-log'' because of logarithmic corrections in correlation functions.} boundary transition happens when $2\leq N \leq N_c$ \cite{Metlitski:2020cqy}. $N_c$ is determined by parameters which can be extracted by studying another boundary universality class, called the normal transition, where an ordering field is applied on the boundary. At the normal transition, the crossing equation (bulk-to-boundary bootstrap equation) for the two-point function of real scalars is schematically
\begin{align}
	\label{eq:bulktoboundaryequ}
	1 + \sum_{k} \lambda_{k} f_{\text{bulk}}(\Delta_{k}, \xi) = \xi^{\Delta_{\phi}} \left( \mu_{\phi} + \sum_{n} \mu_{n} f_{\text{bry}}(\hat{\Delta}_{n}, \xi) \right)
\end{align}
where $\sum_{k}$ is over the bulk operator and $\xi$ is a bulk-boundary cross ratio; $\sum_{n}$ is the sum over the boundary operators. In the case of $O(N)$ CFT, these terms will be dressed with the $O(N)$ tensor structure, which contains $N$ explicitly. Then, certain OPE coefficients determine $N_c$. Specifically, the extraordinary-log phase transition happens when the quantity $\alpha=\frac{1}{32\pi }\frac{\mu _{\sigma }}{\mu _t}-\frac{N-2}{2\pi }$ is positive, where $\mu _t, \mu _{\sigma }$ are certain OPE coefficients.

The OPE coefficients in Eq.~\eqref{eq:bulktoboundaryequ} are, in general, not positive. However, in this specific case, the $2+\epsilon$, large-$N$, $4-\epsilon$ calculation all show these coefficients are positive. Ref.~\cite{Padayasi:2021sik} also studied the system using Gliozzi's method, which show no negativity in those OPE coefficients. The work then assumed the positivity of coefficients and formulate the problem as a SDP. It applied the standard approach to map out the feasible region in $\alpha$ vs. $N$. The work found the bound $N_c>3$, which is rigorous if one accepts the positivity assumptions. With the Gliozzi method, the work found $N_c \approx 4$, consistent with Monte Carlo results. This work is an example where the standard SDP method could be applied to study the bulk-to-boundary bootstrap equation in some situations.

\section{Delaunay triangulation and surface cutting}

It is a natural expectation that incorporating more and more crossing equations should increase the constraining power of the conformal bootstrap. This has been demonstrated in \cite{Kos:2014bka,Kos:2015mba,Kos:2016ysd}, where for the first time more than one crossing equation was used. Furthermore Ref.~\cite{Kos:2016ysd} showed that the non-degeneracy of an \emph{isolated} exchanged operator is a powerful constraint. This isolated non-degenerate operator may be one of the external operators, or an operator which is otherwise known to be non-degenerate. When the non-degeneracy condition is imposed, the contribution of an isolated operator to a crossing equation is proportional to a rank-1 positive-semidefinite matrix, parametrized by the OPE coefficients external-external-exchanged.\footnote{While for non-isolated exchanged operators we have a positive-semidefinite matrix without rank restriction.}  Thus the corresponding SDPs usually depend on two classes of parameters: the scaling dimensions of the external operators and the OPE coefficients external-external-exchanged for the isolated exchanged operators. In typical bootstrap studies, one wants to map out the allowed region (i.e.~the region consistent with the crossing equations) in the parameter space. For every point in the parameter space, {\tt SDPB} can tell us if the point is allowed or not. We would like to perform a scan of the parameter space, running {\tt SDPB} for many points, and then infer the shape of the boundary of the allowed region.\footnote{Here we are describing the so called ``oracle mode'' which was the standard way of running bootstrap computations before the advent of the navigator function, to be described in Section \ref{sec:navigator}.}

As we consider correlators involving more and more external operators, the dimension of the parameter space increases rapidly. Exploring such a high dimensional space is a major challenge for numerical bootstrap study of large correlator systems. A brute force scanning approach would suffer from the ``curse of dimensionality", as the number of SDP runs will increase exponentially with the dimension of space. Here we would like to highlight the algorithms developed in \cite{Chester:2019ifh,Chester:2020iyt}, which partially address this challenge.  Using those algorithms, the cited references achieved remarkable progress on the critical exponents of 3D $O(2)$ and $O(3)$ vector models. We will now review the main ideas and these applications. 

\subsection{Algorithms}
\label{sec:algorithms}

As mentioned above, we must perform a scan in the space of scaling dimensions times OPE coefficients. These two scanning directions are handled separately. For the scaling dimensions, one uses an adaptive sampling method, called \emph{Delaunay triangulation}, to map out the boundary of the allowed region. One first computes a grid of points that contains both allowed and disallowed points, then one applies the Delaunay triangulation \cite{Delauney} which finds a special set of triangles that link those points. The triangles that contain both allowed points and disallowed points are the triangles covering the boundary. The area of those triangles roughly indicates the local resolution of the boundary. One then ranks those triangles by area and samples the middle points of the largest triangles, i.e.~focus on improving the regions with low resolution. This method is essentially a higher dimensional generalization of the 1D bisection method. It works well for 2D and 3D space. This was enough for the applications in \cite{Chester:2019ifh,Chester:2020iyt}. For higher dimensional space, the computations become much slower and the ``curse of the dimensionality" strikes back.

As for the scan in the space of the OPE coefficients, the work \cite{Chester:2019ifh} developed a novel method, called the \emph{cutting surface algorithm}, which is significantly faster than the Delaunay triangulation method. Here we describe the basic idea using a low-dimensional example, and then briefly discuss the higher-dimensional case.

A non-degenerate isolated internal operator gives rise to a term in the crossing equation of the form $\lambda ^T.\vec V.\lambda$ where $\lambda\in \mathbb{R}^n$ is the vector of its OPE coefficients with the external operators and $\vec V$ is a vector of $n\times n$ matrices. Given a trial point $\lambda_1$, we use {\tt SDPB} to look for a functional $\alpha_1$ such that $\lambda_1 ^T.\alpha_1 (\vec V).\lambda_1 \geq 0$. If such a functional exists, the point $\lambda_1$ is ruled out, otherwise one concludes $\lambda_1$ is an allowed point. The key observation is that $\alpha_1$, if it exists, rules out not only $\lambda_1$ but also a sizable disallowed region around $\lambda_1$. If $\lambda$ is a 2D vector, this disallowed region can be easily found explicitly by solving the quadratic inequality $\lambda^T.\alpha_1 (V).\lambda\geq 0$.\footnote{As the condition is invariant under rescalings, it is enough to consider $\lambda=(1,\bar\lambda)$.} The next trial point $\lambda_2$ can now be chosen outside of this disallowed region, and so on. The disallowed region, which is the union of disallowed regions of $\lambda_1,\lambda_2,\ldots$, quickly grows. In the end, one ends up with an allowed point, or the disallowed region covers the entire space.

When $\lambda\in \mathbb{R}^n$, $n\ge 3$, the disallowed region has to be defined implicitly by a set of quadratic inequalities $\lambda^T.\alpha_i (V).\lambda\geq 0$, $i=1,\ldots,k$. One decides the new trial $\lambda_\text{next}$ in step $k+1$ by two steps: (1) try to find a point that is outside the disallowed region, i.e. $\lambda_\text{next}^T.\alpha_i (V).\lambda_\text{next}< 0$ for all $\alpha_i$; (2) try to move this point away from the disallowed region as much as possible. The reason for (2) is that, if the new point $\lambda_\text{next}$ is chosen roughly at the center of the undetermined region, the new functional $\alpha_\text{next}$ (if it exists) usually rules out half of that region. Since each iteration cuts off about half of the volume, the total number of steps to achieve the needed accuracy grows roughly linearly in $n$. This is a much faster performance than for the Delaunay triangulation, achievable thanks to the quadratic structure of the constraints. The step (1) is a type of problem called \textit{quadratically constrained quadratic program} (QCQP), which in NP-hard in general. However for the problem at hand, \cite{Chester:2019ifh} founds several heuristic approaches that work. The heuristics are not rigorous and should not be expected to solve generic QCQPs. What sets our problem apart is that one often has some idea about the expected size of the OPE coefficients, and a bounding box can then be set to constrain the space. With a suitable bounding box, those heuristics work very well, at least when the number of OPE coefficients is not too big. When the heuristics fail to find an allowed point, one declares the OPE coefficient space is ruled out. To speed up the computation, each SDP computation is ``hot-started" \cite{Go:2019lke}, namely one reuses of the final state of the SDP solver from the previous computation as the initial state in the new computation.

To combine Delaunay triangulation with cutting surface, one proceeds as follows. First one chooses a grid of points in the space of scaling dimensions and computes the allowedness of each point, which means using the cutting surface algorithm to rule out (or find allowed) OPE coefficients inside the bounding box (at fixed scaling dimensions). Then one uses the Delaunay triangulation to refine the grid in the space of scaling dimensions. This combined algorithm is suitable for 2 or 3 scaling dimensions, and a few OPE coefficients.\footnote{In Ref. \cite{Chester:2022hzt}, it was used to scan 7 ratios of OPE coefficients. As the number of OPE coefficients grows, one generally has to set a smaller bounding box; otherwise the heuristics for the QCQP might fail.} It is implemented in both \tmtt{hyperion} and \tmtt{simpleboot} frameworks.

\subsection{Application to the $O(2)$ model: the $\nu$ controversy resolved}

The 3d $O(2)$ universality class describes critical phenomena in many physical systems and has been studied intensively both experimentally and theoretically. Experimentally, the most precise measurement of the 3d $O(2)$ critical exponents came from the study of ${}^4$He superfluid phase transition on the Space Shuttle Columbia in 1992 \cite{Lipa:1996zz,Lipa:2000zz,Lipa:2003zz}. The critical exponent $\nu$ obtained from the data analysis of this experiment was $\nu^\text{EXP}  = 0.6709(1)$. 

Theoretically, the simplest description in continuum field theory is given by the Lagrangian, where the mass term $m^2$ needs to be fine-tuned to reach the critical point:
\begin{align}
	\label{eq:ONlag}
	{\cal L } = \frac{1}{2} |\partial \vec \phi|^2  +  \frac{1}{2} m^2 |\vec \phi|^2  + \frac{g}{4!} |\vec \phi|^4\,,
\end{align}
where $\phi$ transforms in the fundamental representation of $O(2)$ and $s=|\vec \phi|^2$ is the leading singlet. Their scaling dimensions are related to critical exponents by
\begin{align}
	\Delta_\phi = \frac{1+\eta}2 \,,\qquad \Delta_s = 3-\frac1{\nu}.
\end{align}
They can be calculated using the renormalization group method. More accurate results however came from the Monte Carlo (MC) simulation of \cite{Hasenbusch:2019jkj}, which estimated $\nu^\text{MC} =0.67169(7)$. Intriguingly, there is a 8$\sigma$ difference between $\nu^\text{MC}$ and $\nu^\text{EXP}$.

The bootstrap study of the 3d $O(N)$ model was initiated in \cite{Kos:2013tga} by considering the correlation function $\<\phi_i\phi_j \phi_k\phi_l\>$. Later, in \cite{Kos:2015mba,Kos:2016ysd}, all correlation functions of $\phi,s$ were bootstrapped, which allowed to constrain the scaling dimensions $\Delta_\phi$, $\Delta_s$ to an island, although the size of that island was not yet sufficiently small to discriminate between $\nu^\text{MC}$ and $\nu^\text{EXP}$. 

\begin{figure}[t!]
\centering
\includegraphics[width=0.45\textwidth]{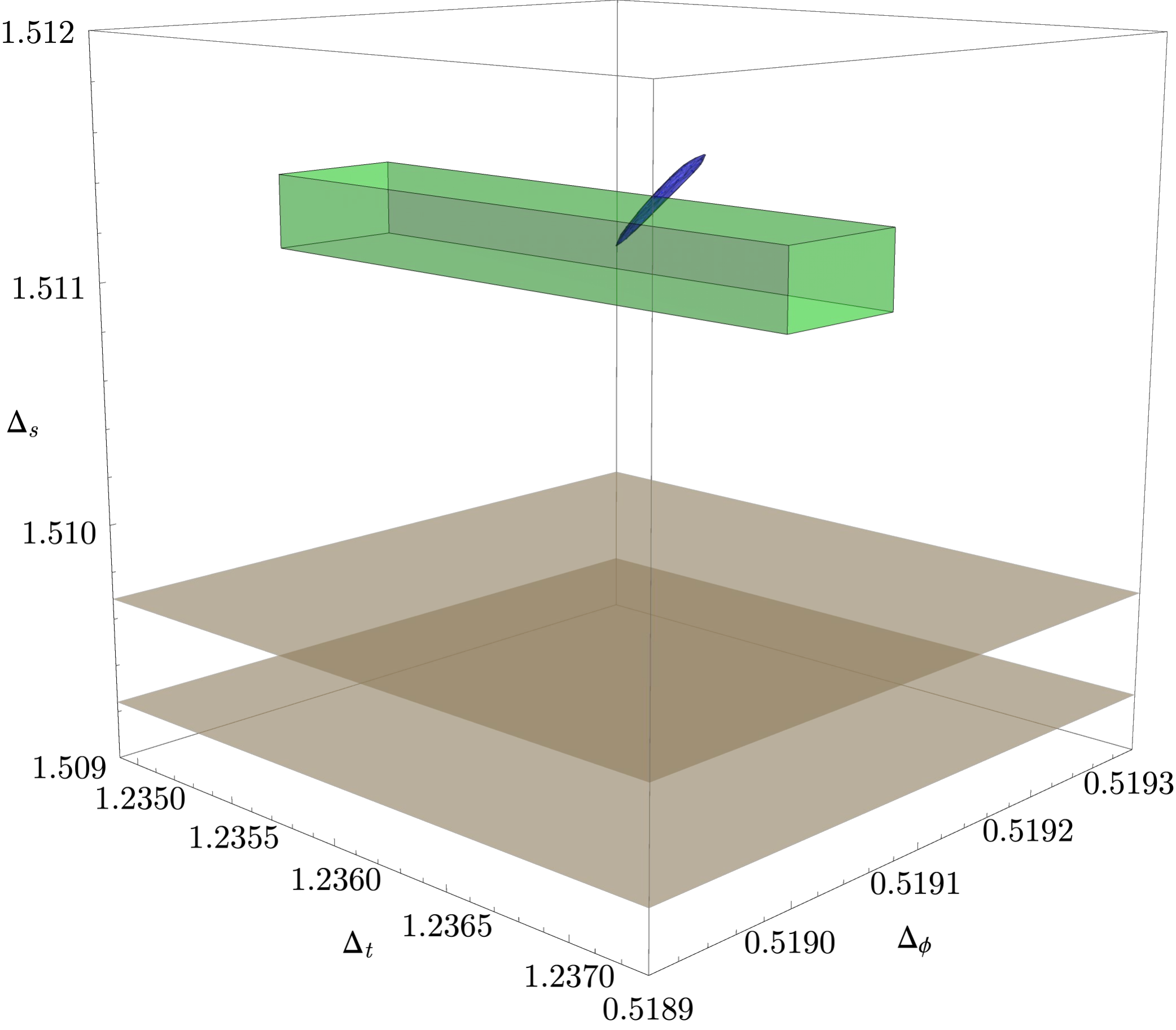}
\caption{\label{fig:O2Island}
(Color online) The blue (dark gray) island is the allowed region from \cite{Chester:2019ifh} for the scaling dimensions of $s,\phi,t$. The green (gray) box indicates results from the Monte Carlo studies \cite{Hasenbusch:2011zwv,Hasenbusch:2019jkj}. The brown (gray) planes represent the $1\sigma$ confidence interval from the experiment \cite{PhysRevB.68.174518}. Figure from \cite{Chester:2019ifh}, \CCBY.
}
\end{figure}

Finally, Ref.~\cite{Chester:2019ifh} considered an even larger system of correlation functions, involving three relevant fields of this model as external operators: scalars $\phi, s$ and the charge-2 operator $t_{i j}=\phi_i \phi_j-\frac{1}{2}\delta_{i j} (\phi_k \phi^k)$. Thanks to employing the cutting surface algorithm and pushing to a high derivative order, this paper could resolve the $\nu^\text{MC}$/$\nu^\text{EXP}$ puzzle, as we now describe.\footnote{The analysis with three external operators was also carried out in \cite{Go:2019lke}. Without cutting surface algorithm, this reference could go only to a relatively low derivative order, not enough to solve the puzzle.} The assumptions of \cite{Chester:2019ifh} on the spectrum are that $\phi,s,t$ are the only relevant operators in the charge-0,1,2 sectors, while the charge-3 scalars having dimension larger than 1\footnote{The leading charge-3 scalar dimension is $\approx 2.1$.} and all charge-4 scalars irrelevant.\footnote{In addition, tiny gap above the unitarity bound were assumed in many sectors. This is totally reasonable since we do not expect operators at unitarity except for the stress tensor and the conserved current. The technical reason for this assumption is to avoid poles of the conformal blocks at the unitarity bound, which interfere with the SDP numerics. See \cite{Chester:2019ifh}, Section 3.6 for a detailed explanation.} These assumptions are physically reasonable. Under these assumptions, \cite{Chester:2019ifh} explored the parameter space spanned by 3 dimensions of $\{\phi,s,t\}$ and 3 OPE coefficient ratios $\{\lambda_{sss}/\lambda_{\phi\phi t}, \lambda_{\phi\phi s}/\lambda_{\phi\phi t}, \lambda_{tts}/\lambda_{\phi\phi t}\}$. The Delaunay triangulation/cutting surface algorithm described above was used to scan over this 6-dimensional space efficiently. The most constraining computation was done at the derivative order $\Lambda=43$, and consumed 1.03 million CPU core-hours. The results are summarized in Table \ref{tab:O2vstresult} and Figure \ref{fig:O2Island}.

\begin{table}[ht]
	\centering
	\begin{tabular}{@{}|c|c|@{}}
		\hline
		CFT data & value \\
		\hline
		$\Delta_s$  & 1.51136({\bf22}) \\
		$\Delta_\phi$ & 0.519088({\bf22}) \\
		$\Delta_t$  &1.23629({\bf11}) \\
		\hline 
	\end{tabular}
	\caption{Results of \cite{Chester:2019ifh} for the scaling dimensions of $s,\phi,t$, the leading charge 0, 1, and 2 scalars. Uncertainties in bold are rigorous.
		\label{tab:O2vstresult}}
\end{table}

The found $\Delta_s$ translates into the critical exponent $\nu$ from the conformal bootstrap: $\nu^\text{CB}=0.671754(\bf 99)$. The uncertainty in bold font is rigorous because it was determined from an allowed bootstrap island, which may only shrink as more and more constraints are added in future studies.\footnote{For completeness, it should be mentioned that there are some other sources of ``error in the error'' which are not fully rigorous, while being under control. For example, conformal block derivatives are replaced by their rational approximations when passing to an SDP. The error from this approximation is monitored, and we estimate its effect to be orders of magnitude smaller than the main error reported above, coming from the size of the bootstrap island. If needed, this error can be easily decreased further. Another non-rigorous error comes from the Delaunay triangulation method and the heuristics in the cutting surface algorithm not being completely rigorous. The study \cite{Chester:2019ifh} took measures to control this issue. For example, the reported error bar has added uncertainty in the Delaunay triangulation method, roughly represented by the size of the triangle on the boundary. These sources of error could be completely removed by repeating the study of \cite{Chester:2019ifh} using the navigator method of Section \ref{sec:navigator}.} This value is consistent with $\nu^\text{MC}$ but decisively rules out the experimental measurement $\nu^\text{EXP}$. It would be interesting to perform a new statistical analysis of the experimental data to understand if the experimental error was perhaps underestimated.

\subsection{Application to the $O(N)$ Gross-Neveu-Yukawa model}
\label{sec:GNY}

The Gross-Neveu-Yukawa (GNY) model with $O(N)$ symmetry is described by the following Lagrangian:
\begin{align}\label{Lag_GNY}
\mathcal{L} = \frac{1}{2} (\partial \phi)^2 + \frac{i}{2} \psi_i \slashed{\partial} \psi_i + \frac{1}{2} m^2 \phi^2 + \frac{\lambda}{2} \phi^4 + i \frac{g}{2} \phi \psi_i \psi_i,
\end{align}
where $\phi$ is a parity-odd scalar, and $\psi$ represents $N$ Majorana fermions transforming in the vector representation of $O(N)$. The $O(N)$ singlet $\phi$ couples to $\psi_i$ through the Yukawa coupling term $\phi \psi_i \psi_i$. At a certain value of the mass $m$, the model becomes critical and flows to a CFT. When $N=1$, the model is the same as (\ref{Lag_GNYsuperIsing}) and is expected to exhibit emergent $\mathcal{N}=1$ supersymmetry.

The large-$N$ expansion of the scaling dimensions of various operators in this model was computed in \cite{Gracey:2018fwq, Manashov:2017rrx}. Based on the large-$N$ results, some important features of the spectrum are: (1) there is only one single relevant operator, $\epsilon \sim \phi^2$; (2) due to the equation of motion, the leading parity-even fermion $\chi_i \sim \phi^3 \psi_i$ has a large scaling dimension, and large-$N$ expansion predicts $\Delta_\chi \ge 4$; (3) although not obvious from the equation of motion, in large-$N$ computations, $\sigma \sim \phi$ is the only relevant parity-odd singlet for $N\ge 2$.


These features are suitable as input for a bootstrap study of the CFT. The work \cite{Erramilli:2022kgp} conducted a bootstrap analysis on all four-point functions of $\sigma, \epsilon, \psi_i$ with reasonable and mild assumptions on $\Delta_{\sigma'},\Delta_{\epsilon'}, \Delta_{\psi'}, \Delta_{\chi'}$, where $\mathcal{O'}$ denotes the next operator in the same sector as $\mathcal{O}$. The conformal block decomposition involving fermions was effectively computed using \texttt{block\_3d} (Section \ref{CB}). 

To impose that $\sigma, \epsilon, \psi$ are the only operators at their scaling dimensions, one must scan over the OPE coefficients involving only external operators. Therefore, the free parameters include the scaling dimensions $\Delta_\sigma, \Delta_\epsilon, \Delta_\psi$ and the OPE coefficients $\lambda_{\psi\psi\sigma},\lambda_{\psi\psi\epsilon},\lambda_{\sigma\sigma\epsilon},\lambda_{\epsilon\epsilon\epsilon}$. Similar to \cite{Chester:2019ifh}, Ref.~\cite{Erramilli:2022kgp} applied the Delaunay search algorithm to scan over the scaling dimensions, and the cutting-surface algorithm to scan over the three ratios of OPE coefficients. These techniques sufficed to get very interesting results which we will now describe.

For $N=1$, \cite{Erramilli:2022kgp} assumed $\Delta_{\sigma'} > 2.5, \Delta_{\epsilon'} > 3, \Delta_{\psi'} > 2, \Delta_{\chi'} > 3.5$ and found an isolated region. The super-Ising island of \cite{Atanasov:2022bpi} is located at the tip of this isolated region. Without assuming supersymmetry, \cite{Erramilli:2022kgp} checked whether a conserved supercurrent exists in the isolated region. Near the tip, corresponding to the range of parameters in Table \ref{tab:superIsingresult}, the spin-3/2 upper bound was found to be $\Delta_{\text{SC}}<2.5003219$, very close to the exactly conserved supercurrent dimension 2.5. This implies that any CFT with these parameters must be, to an extremely high degree of precision, supersymmetric — strong evidence for the emergent supersymmetry scenario.

\begin{figure}[t!]
	\centering
	\includegraphics[width=.48\textwidth]{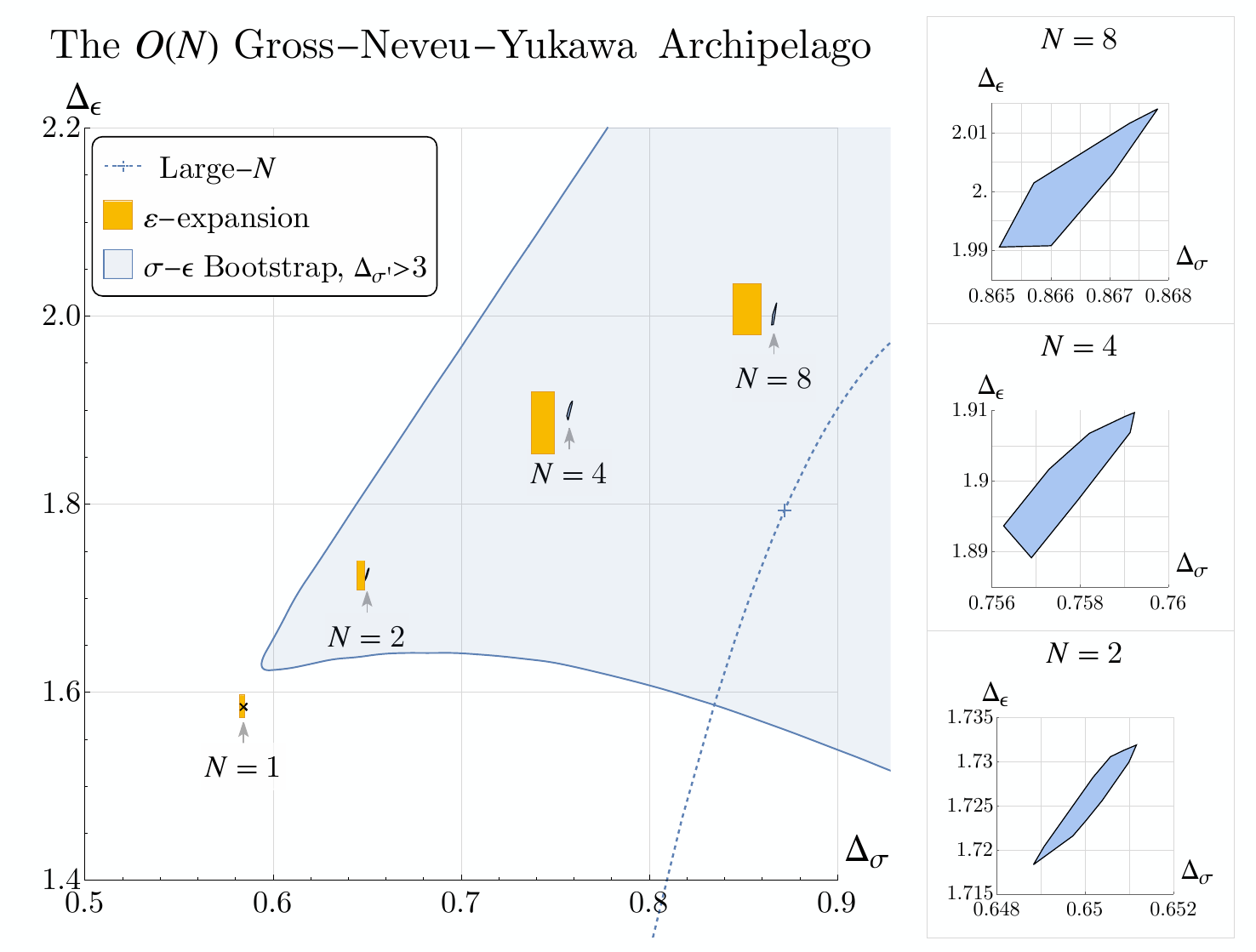}
	\caption{(Color online) Right: zoomed-in bootstrap islands for $N=2,4,8$ $O(N)$ GNY model from \cite{Erramilli:2022kgp} at $\Lambda=35$, projected onto the \((\Delta_\sigma,\Delta_\epsilon)\) plane. Left: zoomed-out view of the same islands. Dotted blue curve: perturbative estimates in the large-\(N\) expansion. Orange (dark gray) boxes: Borel-resummations of the \((4-\epsilon)\)-expansion \cite{Ihrig:2018hho}. The ``x" indicates the location of the \(N=1\) island of \cite{Atanasov:2022bpi}. Light blue (light gray) region: the general \(\sigma\)-\(\eps\) bootstrap bounds with the assumption \(\Delta_{\sigma'}>3\) from \cite{Atanasov:2018kqw}.  Bootstrap islands do not overlap with the orange (dark gray) boxes for $N=2,4,8$, implying that the error bar of the Borel-resumed results was underestimated. Figure from \cite{Erramilli:2022kgp}, \CCBY, colors modified. \label{fig:archipelago-sig-eps-intro}}
\end{figure}

For $N=2, 4, 8$, \cite{Erramilli:2022kgp} assumed $\Delta_{\sigma'} > 3, \Delta_{\epsilon'} > 3, \Delta_{\psi'} > 2, \Delta_{\chi'} > 3.5$ and found small bootstrap islands for those theories. Figure \ref{fig:archipelago-sig-eps-intro} shows these islands at $\Lambda=35$, and the scaling dimensions with rigorous error bars are summarized in Table \ref{tab:bootstrap-GNY-summary}.

\begin{table}[htp]
	\begin{center}
		\resizebox{\columnwidth}{!}
		{
			\begin{tabular}{@{}lclclclclclcl@{}}
				\hline
				& \hspace{0.5cm} $\Delta_\psi$ & \hspace{0.5cm} $\Delta_\sigma$ &\hspace{0.5cm}  $\Delta_\epsilon$ &\hspace{0.5cm}  $\eta_\psi$ &\hspace{0.5cm}  $\eta_\phi$ &\hspace{0.5cm}  $\nu^{-1}$ \\
				\hline
				$N=2$
				& $1.06861\mathbf{(12)}$ & $0.6500\mathbf{(12)}$ & $1.725\mathbf{(7)}$ & $0.13722\mathbf{(24)}$ & $0.3000\mathbf{(23)}$ & $1.275\mathbf{(7)}$\\
				$N=4$
				& $1.04356\mathbf{(16)}$& $0.7578\mathbf{(15)}$ & $1.899\mathbf{(10)}$ & $0.08712\mathbf{(32)}$ & $0.5155\mathbf{(30)}$ & $1.101\mathbf{(10)}$  \\
				$N=8$
				& $1.02119\mathbf{(5)}$ & $0.8665\mathbf{(13)}$ & $2.002\mathbf{(12)}$ & $0.04238\mathbf{(11)}$ & $0.7329\mathbf{(27)}$ & $0.998\mathbf{(12)}$ \\ 
             \hline
			\end{tabular}
		}
		\caption{Scaling dimensions and critical exponents obtained in for the $O(N)$ GNY model, with rigorous error bars in boldface \cite{Erramilli:2022kgp} .
			\label{tab:bootstrap-GNY-summary}
		}
	\end{center}
\end{table}

It should be noted that there is another closely related Lagrangian
\begin{align}\label{Lag_GNY_generic}
\mathcal{L} &= \frac{1}{2} (\partial \phi)^2 + \frac{i}{2} \psi_i^A \slashed{\partial} \psi_i^A + \frac{1}{2} m^2 \phi^2 + \frac{\lambda}{2} \phi^4 \nonumber \\ 
&+ i\frac{ g_1}{2} \phi (\psi_i^L \psi_i^L - \psi_i^R \psi_i^R) + i\frac{ g_2}{2} \phi (\psi_i^L \psi_i^L + \psi_i^R \psi_i^R),
\end{align}
where \(i = 1, \dots, \frac{N}{2}\) is an \(O(N/2)\) vector index and \(A = L, R\) labels two species of fermions. 
If \(g_1 = 0, g_2 \neq 0\), it corresponds to the \(O(N)\) GNY Lagrangian (\ref{Lag_GNY}), and its fixed point is bootstrapped in \cite{Erramilli:2022kgp}. If \(g_1 \neq 0, g_2 = 0\), the model has \(O(N/2)^2 \rtimes Z_2\) symmetry, where \(Z_2\) is realized by \(\psi_i^L \leftrightarrow \psi_i^R\), \(\phi \to -\phi\). At a critical value of mass, the \(O(N/2)^2 \rtimes Z_2\) GNY model is expected to flow to the ``chiral Ising" universality class, which is different from the \(O(N)\) GNY CFT. Various Monte Carlo simulations \cite{Huffman:2019efk, Liu:2019xnb} have studied this phase transition.\footnote{We thank Yin-Chen He for clarifying this point.} There are interesting applications of this model in graphene and D-wave superconductors (see \cite{Erramilli:2022kgp} for a summary). It would be interesting to see if future bootstrap studies of the \(O(N/2)^2 \rtimes Z_2\) GNY CFT can distinguish its critical exponents from those of the \(O(N)\) GNY CFT.

\section{Tiptop}

In many situations, we are interested in the maximum or minimum value of a certain parameter over the allowed region. The {\tt tiptop} algorithm \cite{Chester:2020iyt} was designed to explore the ``tip'' of a convex region. Like Delaunay triangulation, {\tt tiptop} usually works in pair the cutting surface algorithm: {\tt tiptop} recommends new points near the tip in a direction of a certain scaling dimension, while the cutting surface algorithm is called to decide the (dis)allowedness of those points. In this section we briefly describe this algorithm and its application to the 3D $O(3)$ vector model. The algorithm was also used in \cite{Mitchell:2024hix} to bound the leading irrelevant operator of the 3d Gross-Neveu-Yukawa CFTs from Section \ref{sec:GNY}.

\subsection{The tiptop algorithm}

We consider a bootstrap problem depending on $n+1$ parameters: $x\in \mathbb{R}^n$ and $y$. We want to know the maximum value of $y$ in the allowed region. For example, $x$ may comprise $n$ scaling dimensions and $y$ another scaling dimension we are particularly interested in. We assume the allowed region around the maximum $y$ is convex, so that as $y$ increases the allowed region of $x$ (for a fixed $y$) shrinks to zero size. The {\tt tiptop} algorithm starts with some disallowed points and at least one allowed point at current $y=y_\text{allowed}$. Given a list of allowed, disallowed, and
in-progress points (which are treated as disallowed by {\tt tiptop}), it recommends one point to look at next. 

To do that, {\tt tiptop} first checks if the shape and the size of the allowed region at $y_\text{allowed}$ are well understood. An affine coordinate transformation of the $x$ space is performed, such that the region of known allowed $x$ at $y_\text{allowed}$ is roughly spherical.\footnote{The affine transformation is needed, because isolated allowed regions in conformal bootstrap tend to have extreme aspect ratios.} 
The full region of interest is then rescaled so that it is the cube $[-1,1]^n$. One recursively subdivides this region into \emph{cells}, which are cubes of size $2^{-k}$,  $k=0,1,\ldots, \min(K,k_{\rm max})$ , where $K$ is the first integer such that $2^{-K}$ is less than $f$ times the minimum coordinate extents of the set of allowed points, $f$ is a user-defined parameter (e.g.~$f=2$ works well), and $k_{\rm max}=47$.\footnote{The value of $k_{\rm max}$ was chosen so that $2^{-k_{\rm max}}$ is somewhat larger than the minimum resolution of an IEEE-754 double-precision number.}  The algorithm then recommends a new point to be placed in the largest empty cell (i.e.~a cell in which there is no point), which is diagonally adjacent to a cell containing allowed points. 
	
If there is no such empty cell, the shape of allowed region at $y_\text{allowed}$ is considered well-understood. In this case, the algorithm recommends a new point at a higher $y$. The $x$ of the new point is roughly in the center of allowed points at $y_\text{allowed}$, while $y$ is half-way between $y_\text{allowed}$ and $y_\text{ceiling}$ (i.e.~a bisection step). At first, $y_\text{ceiling}$ is a user-defined value safely larger than $y_{\max}$. After some points have been checked, $y_\text{ceiling}$ is the lowest disallowed $y$. If $y_\text{allowed}-y_\text{ceiling}$ is smaller than a certain user-specified value, the algorithm stops. 

\subsection{Application to the $O(3)$ instability problem}

The $O(3)$ vector model can be defined by the Lagrangian \eqref{eq:ONlag} with $\phi$ in the vector representation of $O(3)$. Alternatively, the Heisenberg model on the cubic lattice with the Hamiltonian 
\begin{align}
	\label{eq:HeisenbergHamiltonian}
	{\cal H } = -J\sum _{\langle x,y\rangle }\sigma_x \cdot \sigma_y,
\end{align}
where $\sigma_x\in \mathbb{R}^3$ are classical spins of unit length, has the phase transition in the same universality class.

The work \cite{Chester:2020iyt} bootstrapped correlators of $\phi,s,t$, which are the leading scalars in the $O(3)$ vector, singlet, rank-2 traceless symmetric tensor irreps. The bootstrap setup of the $O(3)$ CFT is similar to the $O(2)$ case except that there are more irreducible representation channels in the OPE expansion $t\times t$. Of particular interest is the operator $t_4$, the leading 4-index traceless symmetric tensor, appearing in this OPE.

With the assumption that $\phi,s,t$ are the only relevant scalars in their representation, the work \cite{Chester:2020iyt} made two bootstrap computations. In the first computation, a conservative assumption $\Delta_{t_4}\ge 2$ was made. Then, Delaunay triangulation and cutting surface were used to map out the $O(3)$ allowed island at the derivative order $\Lambda=43$. Rigorous results for the scaling dimensions of $s,\phi,t$ from this computation are given in Table \ref{tab:O3vstresult}.

\begin{figure}[t!]
	\centering
	\includegraphics[width=0.49\textwidth]{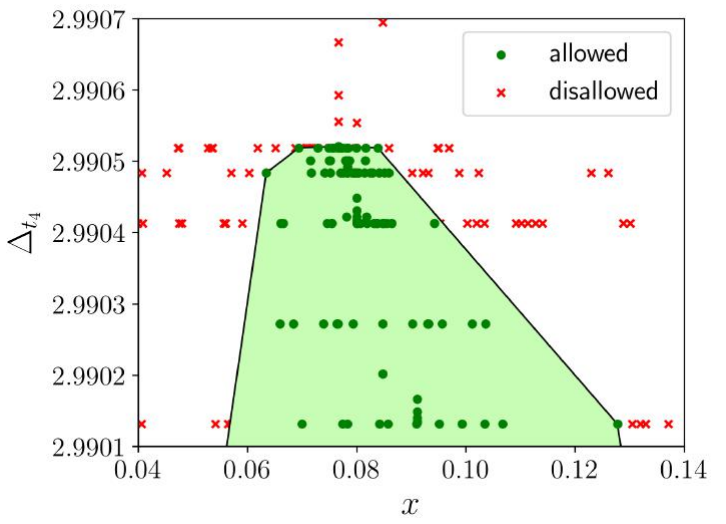}
	\caption{\label{fig:tiptop} (Color online) A two-dimensional projection of the progress of {\tt tiptop} as it was maximizing $\Delta_{t_4}$. Figure from \cite{Chester:2020iyt}.}
\end{figure}

\begin{table}[ht]
	\centering
	\begin{tabular}{@{}|c|c|@{}}
		\hline
		CFT data & value \\
		\hline
		$\Delta_s$  & 1.59488({\bf81}) \\
		$\Delta_\phi$ & 1.518936({\bf61}) \\
		$\Delta_t$  &1.20954({\bf32}) \\
		\hline 
	\end{tabular}
	\caption{Conformal bootstrap results of \cite{Chester:2020iyt} for the scaling dimensions of $s,\phi,t$, the leading scalars in the $O(3)$ vector, singlet, rank-2 traceless symmetric tensor irreps. Bold uncertainties are rigorous.
		\label{tab:O3vstresult}}
\end{table}
In the second computation, \cite{Chester:2020iyt} used the {\tt tiptop} algorithm to find a rigorous upper bound on $\Delta_{t_4}$, with the result $\Delta_{t_4}\le 2.99056$ at $\Lambda=35$. Therefore $t_4$ is relevant, although very weakly so. Fig.~\ref{fig:tiptop} gives an idea of the progress of {\tt tiptop} as it was maximizing $\Delta_{t_4}$.

The relevance of $t_4$ is important for the structure of the RG flow.
Since $t_4$ is relevant, the Heisenberg fixed point is unstable with respect to perturbations by (a linear combination of components of) $t_4$. A particularly interesting linear combination is 
\begin{equation}
	\label{eq:cubic}
	\sum_{i=1}^3 (t_4)_{iiii},
\end{equation}
which preserves the symmetry group $B_3 \equiv S_3\ltimes (\mathbb{Z}_2)^3 \subset O(3) $, called the cubic group.\footnote{The $S_3$ permutes the three coordinate axes, and the $(\mathbb{Z}_2)^3$ flips those three axes.} This triggers an RG flow to another fixed point called the cubic fixed point \cite{PhysRevB.8.4270}, whose symmetry group is $B_3$. Since $t_4$ is so weakly relevant, this RG flow is extremely short, and the critical exponents of the Heisenberg and cubic fixed point are very close to each other. 

Previously, the (ir)relevance of $t_4$ was studied for many decades in perturbation theory.
These studies compute the critical exponent $Y=3-\Delta_{t_4}$. Since $Y$ is close to zero, it is not easy to determine its sign. By year 2000, RG studies converged to the conclusion that $Y_{O(3)}>0$ at the Heisenberg fixed point, while $Y_{B_3}<0$ at the cubic fixed point (see \cite{Pelissetto:2000ek}, Sec.~11.3, Table 33). However the error bars of these studies were significant, and the sign of $Y$ was only determined at a 2$\sigma$ level. Monte Carlo simulations improved this to 3$\sigma$ in 2011 \cite{Hasenbusch2011}. The above bootstrap result gives a rigorous proof showing that $t_4$ is indeed relevant.\footnote{Since then, Hasenbusch obtained an accurate Monte Carlo determination $Y_{O(3)} = 0.0142(6)$, $Y_{B_3}=-0.0133(8)$ \cite{Hasenbusch:2022zur}, while a conformal bootstrap calculation with $v$, $s$, $t_2$, $t_4$ external scalars \cite{Rong:2023owx} computed the OPE coefficients of $t_4$ operator and set up a conformal perturbation theory computation predicting the cubic theory exponents in terms of the $O(3)$ ones.}
\begin{figure}[h]
	\begin{center}
		\includegraphics[scale=1.5]{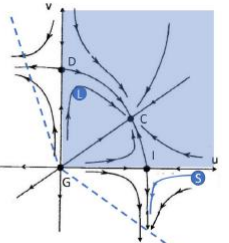}
	\end{center}
	\caption{\label{fig:Aharony}
		(Color online) RG flow in the coupling plane $(u,v)$ for the case $t_4$ is irrelevant at the Heisenberg fixed point (I), driving the flow to the cubic fixed point (C), located at $v>0$. Figure from \cite{Aharony:2022ajv}. The distance between $C$ and $I$ is exaggerated - in reality these fixed points are very close to each other. Also shown in this diagram is the schematic position of $\text{SrTiO}_3$ (S), flowing to a first-order transition, and $\text{LaAlO}_3$ (L), which is attracted to the cubic fixed point and should show a second-order transition with significant corrections to scaling.
	}
\end{figure}

Let us discuss phenomenological implications of the relevance of $t_4$. For ferromagnets this is not so important - the perturbing cubic coupling \eqref{eq:cubic} will have for them a very small coefficient, having spin-orbit origin. An example when this term is important is the structural phase transition in perovskites. Perovskites like $\text{SrTiO}_3$ and $\text{LaAlO}_3$ have crystal cells preserving cubic symmetry at high temperature. As temperature decreases below a certain critical temperature $T_c$, the materials undergo a structural phase transition, where the lattice is stretched in a direction along an axis or a diagonal. Using the Landau theory, this phase transition can be modeled by the potential $\mu  |\vec \phi|^2  + u |\vec \phi|^4 + v \sum_{i=1}^3(\phi_i)^4 $, where $v>0$ and $v<0$ correspond to order parameter $\phi$ breaking the cubic symmetry along an axis or a diagonal.\footnote{Plotting the potential with $\mu<0, v>0$, one can see the minimum of the potential is along the diagonals, and with $\mu<0, v<0$ the minimum is along the axes.} RG studies show that the (stable as we now know for sure) cubic fixed point lies at $v>0$. The RG flow diagram is as in Fig.~\ref{fig:Aharony}.
It follows that perovskites like $\text{SrTiO}_3$, whose low-$T$ structure breaks the cubic symmetry along an axis, will have a first-order phase transition, while $\text{LaAlO}_3$ whose structure breaks cubic symmetry along a diagonal will have a second-order phase transition, in the cubic universality class \cite{Aharony:2022ajv}. However, since $Y$ is so small, the flow is attracted to the cubic fixed point very slowly along the $v$ direction. Hence we expect strong corrections to scaling.

\section{Navigator function}
\label{sec:navigator}

\subsection{General idea}

Delaunay triangulation, surface cutting and {\tt tiptop} algorithms from the
previous sections alleviate the curse of dimensionality in determining the
shape of the allowed region. All these algorithms use {\tt SDPB} in the
oracle mode, testing individual points for being allowed or disallowed. The
navigator function method {\cite{NingSu-letter,Reehorst:2021ykw}} is a radically new idea departing from the
oracle philosophy. In this method a single {\tt SDPB} run returns not a 0/1
information for allowed/disallowed, but a real number whose sign indicates
allowed/disallowed, while the magnitude shows how far the tested point is from
the allowed region boundary. This leads to even more efficient strategies for
multi-parameter bootstrap studies.

As usual, we consider a bootstrap problem characterized by a finite vector of
parameters $x$. Typically, $x$ includes scaling dimensions of a few operators,
their OPE coefficients, and spectrum gap assumptions. Spacetime dimension $d$
and the global symmetry group parameters (such as $N$ for the $O (N)$
symmetry) may also be included in $x$.

\begin{figure}[h]
	\centering
	\includegraphics{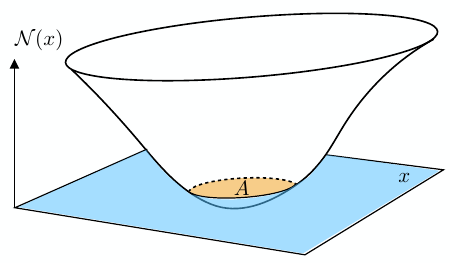}
	\caption{\label{Nschematic}	(Color online) Schematic view of the navigator function $\mathcal{N} (x)$.
		The allowed region $A = \{ x :\mathcal{N} (x) \leqslant 0 \}$.}
\end{figure}

The navigator function {\cite{Reehorst:2021ykw}} is a continuous
differentiable function $\mathcal{N} (x)$ whose sublevel set $\{ x : \mathcal{N} (x) \leqslant 0
\}$ coincides with the allowed region $A$.  This function, in general, is not convex. However, in many examples which have been considered, the function was found to have a nice convex shape in the neighborhood of an isolated allowed region (see Appendix \ref{app:comment}), with a single minimum inside it, see Fig.~\ref{Nschematic}. This property translates into the fact that isolated allowed regions in the conformal bootstrap often have elliptic shapes, which shrink with the increase of the derivative order $\Lambda$. We will see below how to set up such a navigator
function and compute it using {\tt SDPB}. Importantly, the gradient
$\nabla \mathcal{N} (x)$ is also inexpensive to compute \cite{Reehorst:2021ykw}. This is because the navigator will be given as a result of an optimization problem, and first-order variations of the objective at extremality can be found without computing the change in the minimizer.\footnote{This is true even for constrained minimization, as is our case, when a primal-dual optimization method is used such as {\tt SDPB}.}

A typical task from which any bootstrap study starts is to find a single
allowed point. This may be nontrivial if one works at a high derivative order,
so that the allowed region is very small. With the navigator function, an
allowed point is searched for by minimizing $\mathcal{N} (x)$ via a quasi-Newton method
{\cite{Reehorst:2021ykw}} such as BFGS {\cite{wiki:BFGS}}. We start from an
initial guess $x_0$ in the excluded region, and go through a sequence of
points $x_0 \rightarrow x_1 \rightarrow x_2 \rightarrow \cdots .$ If the
allowed region is not empty, we will reach it after a finite number of steps.
If, on the contrary, we reach a positive minimum of $\mathcal{N}(x)$, we conclude that
the allowed region is empty.

Another typical task is to understand the shape of the allowed region. The boundary is the zero set of the navigator $\{ x : \mathcal{N} (x) = 0 \}$. In the case of a 2D allowed region, when the boundary is a curve, we can
fully trace it out by making small steps tangential to the boundary and then projecting back
to the boundary, taking advantage of the available $\nabla \mathcal{N} (x)$. For a higher-dimensional allowed region, we can try to understand the shape of the boundary by understanding its 1D curvilinear sections by various hyperplanes. However, this may be expensive. On the positive side, we are often interested in extremal allowed values in certain directions rather than in the precise shape of the allowed region. We can move inside the allowed region until we hit an extremal point on the boundary in a certain direction, as was done in {\cite{Reehorst:2021ykw,ono2}}. This strategy supersedes the {\tt tiptop} algorithm.

\subsection{Existence and gradient}

It is not trivial that the navigator function exists. We will explain why this
is so on the prototypical example of a single correlator crossing equation:
\begin{equation}
	\sum_{\mathcal{O} \in \phi \times \phi} p_{\phi \phi \mathcal{O}}^{}
	F_{\Delta, \ell} (u, v) = 0,\quad p_{\phi \phi \mathcal{O}} \geqslant 0\,.
	\label{cross}
\end{equation}
We impose the constraints that
\begin{equation}
	\ell = 0 : \qquad \Delta = 0\ (\text{unit operator})\quad {\rm or}\quad \Delta
	\geqslant \Delta_{\ast},
\end{equation}
while for $\ell > 0$ all $\Delta$ must be at or above the unitarity bound
$\Delta_{\ell}$. Eq.~{\eqref{cross}} with these constraints is the primal
problem, depending on $x = (\Delta_{\phi}, \Delta_{\ast})$. The dual problem
is obtained by considering a linear functional $\alpha$ such that:
\begin{align}
	&\alpha (F_{\Delta = 0, \ell = 0} (u, v)) > 0\,,\nn\\
	&\alpha (F_{\Delta, \ell = 0} (u, v)) \geqslant 0\quad {\rm for}\quad \Delta \geqslant
	\Delta_{\ast}\,,\\
	&\alpha (F_{\Delta, \ell} (u, v)) \geqslant 0\quad {\rm for}\quad\Delta \geqslant
	\Delta_{\ell}\,.\nn
\end{align}
As usual, if $\alpha$ solving the dual problem exists, then there is no
$p_{\phi \phi \mathcal{O}} \geqslant 0$ solving the primal problem, and $x$
is disallowed.

The navigator function is defined by solving the following modified dual
problem with an objective:
\begin{align}
	\mathcal{N} (x) = &\max_{\alpha} \alpha (F_{\Delta = 0, \ell = 0} (u, v))\,,\qquad\text{where}\\
	&\alpha (F_{\Delta, \ell = 0} (u, v)) \geqslant 0\quad {\rm for}\quad \Delta \geqslant
	\Delta_{\ast}\,,\nn\\
	&\alpha (F_{\Delta, \ell} (u, v)) \geqslant 0\quad {\rm for}\quad\Delta \geqslant
	\Delta_{\ell}\,,\nn\\
	&	\alpha (F_{{\rm norm}} (u, v)) = 1\,.\nn
\end{align}
Clearly, this problem is of the form which can be handled by {\tt SDPB}. The normalization
vector $F_{{\rm norm}} (u, v)$ should be chosen appropriately to ensure that the navigator is finite.
The criterion for this choice is that the $F_{{\rm norm}} (u, v)$ should lie
strictly inside the cone $\mathcal{C}$ which is generated by all vectors $F_{\Delta, \ell} (u, v)$ allowed to
appear in the crossing equation {\cite{NingL4}}. This naturally leads to the
first navigator construction from {\cite{Reehorst:2021ykw}}, the
$\Sigma$-navigator, when one chooses:
\begin{equation}
	F_{{\rm norm}} (u, v) = \sum_i F_{\Delta_i, \ell_i} (u, v)
\end{equation}
for a finite set of $\Delta_i, \ell_i$ satisfying the spectrum constraints.

In the primal formulation, one computes the navigator as:
\begin{gather}
	\mathcal{N} (x) = \min \lambda\,,\qquad\text{where}\\
	\sum_{\mathcal{O} \in \phi \times \phi} p_{\phi \phi \mathcal{O}}
	F_{\Delta, \ell} (u, v) - \lambda F_{{\rm norm}} (u, v) = 0,\quad p_{\phi \phi
		\mathcal{O}}^{} \geqslant 0\,.\nn
\end{gather}
This leads to the second navigator construction from
{\cite{Reehorst:2021ykw}}, the GFF-navigator, when one chooses $-
F_{{\rm norm}} (u, v)$ to be the sum of a few conformal blocks of a
generalized free field (GFF) solution to crossing. The GFF navigator is
naturally bounded from above by 1.

\subsection{Applications}

Although very recent, the navigator function method has already been applied
in several studies which would be very hard or impossible with previous
techniques
{\cite{Reehorst:2021hmp,Sirois:2022vth,Henriksson:2022gpa,Chester:2022hzt}}.
We would like to give here a brief description of these results.

\subsubsection{Rigorous bounds on irrelevant operators of the 3D Ising CFT }

Reehorst {\cite{Reehorst:2021hmp}} studied the 3D Ising model CFT using the
navigator function depending on 13 parameters: dimensions of 5 operators
$\sigma$, $\epsilon$, $\sigma'$, $\epsilon'$, $T'$ (which is the first spin-2
operator after the stress tensor), central charge $c_T$, and 7 OPE
coefficients $\lambda_{\sigma \sigma \varepsilon}$, $\lambda_{\varepsilon
	\varepsilon \varepsilon}$, $\lambda_{\sigma \sigma \varepsilon'}$,
$\lambda_{\varepsilon \varepsilon \varepsilon'}$, $\lambda_{\sigma \varepsilon
	\sigma'}$, $\lambda_{\sigma \sigma T'}$, $\lambda_{\varepsilon \varepsilon
	T'}$. Imposing gaps $\Delta_{\sigma''}, \Delta_{\varepsilon''}, \Delta_{T''}
\geqslant 6$, he determined an allowed region for the navigator function
parameters. This led to rigorous two-sided bounds on all these parameters
({\cite{Reehorst:2021hmp}},Table 1), for example:
\begin{equation}
	\Delta_{\varepsilon'} = 3.82951 (\mathbf{61}) \qquad \text{} (\Lambda =
	31) \label{eps1rig} .
\end{equation}
Previously, only four of these parameters namely $\Delta_{\sigma},
\Delta_{\varepsilon}, \lambda_{\sigma \sigma \varepsilon}$,
$\lambda_{\varepsilon \varepsilon \varepsilon}$ {\cite{Kos:2016ysd}} had such
rigorous bounds. The bounds on other quantities were determined non-rigorously
{\cite{Simmons-Duffin:2016wlq}} by performing a partial scan over 20 points in
the allowed island in the $\Delta_{\sigma}, \Delta_{\varepsilon},
\lambda_{\sigma \sigma \varepsilon} / \lambda_{\varepsilon \varepsilon
	\varepsilon}$ space, minimizing $c_T$ for each of them, extracting the
spectrum via the extremal functional method {\cite{ElShowk:2012hu}}, and
estimating errors as one standard deviation. For $\varepsilon'$ this
non-rigorous determination gave
\begin{equation}
	\Delta_{\varepsilon'} = 3.82968 (23) \qquad \left(
	\text{{\cite{Simmons-Duffin:2016wlq}}, } \Lambda = 43 \right) .
\end{equation}
We see that the rigorous determination {\eqref{eps1rig}}, though consistent,
has a larger error because it uses smaller $\Lambda$ and because the
non-rigorous method is based on a very partial scan, which may further
underestimate the error. Surprisingly though, for some quantities the rigorous
error turns out to be somewhat smaller than the non-rigorous one, despite
smaller $\Lambda$, e.g.
\begin{equation}
	\lambda_{\varepsilon \varepsilon \varepsilon'} = \left\{ \begin{array}{ll}
		1.5362 (\mathbf{12}) & \left( \text{{\cite{Reehorst:2021hmp}}, } \Lambda
		= 19 \right),\\
		1.5360 (16) & \left( \text{{\cite{Simmons-Duffin:2016wlq}}, } \Lambda = 43
		\right) .
	\end{array} \right.
\end{equation}
As explained in {\cite{Reehorst:2021hmp}}, in this case the non-rigorous
$\lambda_{\varepsilon \varepsilon \varepsilon'}$ determination is polluted by
outlier solutions which contain not one but two nearly degenerate
operators near $\Delta_{\varepsilon'}$, sharing the OPE coefficient
$\lambda_{\varepsilon \varepsilon \varepsilon'}$ (the sharing effect
{\cite{Liu:2020tpf}}). In the navigator function method of
{\cite{Reehorst:2021hmp}}, the operator $\varepsilon'$ is isolated by
definition, excluding the sharing effect and leading to a more robust
determination of $\lambda_{\varepsilon \varepsilon \varepsilon'}$.

\subsubsection{Navigating through the $O (N)$ archipelago}

Previously, Ref. {\cite{Kos:2015mba}} studied the $O (N)$ model in $d = 3$ for
discrete integer values of $N = 1, 2, 3$, {\ldots}, using the scan method.
Using the three-correlators setup $\langle \phi \phi \phi \phi \rangle$,
$\langle \phi \phi s s \rangle$, $\langle s s s s \rangle$, where $\phi, s$
are the lowest scalars in the $O (N)$ vector and singlet irreps, they isolated
the $O (N)$ model in $d = 3$ to islands, referred to as the $O (N)$
archipelago. 

Recently, Sirois {\cite{Sirois:2022vth}} applied the navigator
method to the $O (N)$ model in the same three-correlators setup, but making
$N$ and $d$ to vary continuously. The navigator function depended on four arguments $\Delta_{\phi}, \Delta_s,
\lambda_{s s s} / \lambda_{\phi \phi s}, \Delta_t$, where $t$ is the lowest
scalar in the $O (N)$ symmetric traceless tensor irrep. A gap up to $\Delta =
d$ was imposed in these three channels, as well as gaps of $0.5$ above the
stress tensor and the conserved current operators. With these assumptions, at
$\Lambda = 19$, Ref.~{\cite{Sirois:2022vth}} followed the islands along three
continuous families in the $(d, N)$ space:\footnote{For the first family the navigator function
depended on only 3 arguments $\Delta_{\phi}, \Delta_s, \Delta_t$.}
\begin{eqnarray}
	& \left\{ 1 \leqslant N \leqslant 3, \quad d = 3 \right\}, \nonumber\\
	& \left\{ N = 2, \quad 3 \leqslant d \leqslant 4 \right\}, & 
	\label{eq:3curves}\\
	& \left\{ N = 3, \quad 3 \leqslant d \leqslant 4 \right\} . &  \nonumber
\end{eqnarray}
The position and size of the islands were determined along each line and
compared to the predictions from the $\epsilon$-expansion, finding good
agreement.

This study adds further evidence that one should not be afraid to apply the
unitary numerical conformal bootstrap method to models with noninteger $N$ and
$d$ (for prior evidence in non-integer $d$ see
{\cite{El-Showk:2013nia,Cappelli:2018vir,Bonanno:2022ztf}}). Indeed, while
such models are nominally non-unitary
{\cite{Maldacena:2011jn,Hogervorst:2014rta,Hogervorst:2015akt}}, unitarity
violations are secluded at very high operator dimensions, and are invisible at
the currently attainable numerical accuracy. This should be true as long as
$N$ and $d$ are sufficiently large.

The situation changes however when $N$ becomes too small.\footnote{And also
	when $d$ is too small {\cite{Golden:2014oqa}}.} For example, the limit $N
\rightarrow 1$ is dangerous because the symmetric traceless and antisymmetric
irreps, used in the $O (N)$ model bootstrap analysis, do not exist for $N = 1$
(their dimension goes to zero as $N \rightarrow 1$). Ref.
{\cite{Sirois:2022vth}} found that the island shrunk to zero size when $N
\rightarrow 1^+$ as $d = 3$ (the first family in {\eqref{eq:3curves}}). This
is because there are primary operators in the spectrum at $N > 1$ whose
squared OPE coefficients go linearly to zero as $N \rightarrow 1^+$,
preventing continuation of the unitary solution to crossing to $N < 1$.
Numerically, Ref. {\cite{Sirois:2022vth}} identified two such operators in the
solution to crossing at $d = 3$. Analytically, similar phenomena were shown to
occur in the free $O (N)$ theory, and in the perturvative setting of $d = 4 -
\varepsilon$.

\subsubsection{Ising CFT as a function of $d$: spectrum continuity and level
	repulsion}

Previosuly, the Ising CFT was studied as a function of $d \in [2, 4]$ using
the single-correlator setup $\langle \sigma \sigma \sigma \sigma \rangle$,
identifying the position of the theory with the kink in
$\Delta_{\varepsilon}$-maximization {\cite{El-Showk:2013nia}} or in
the $c_T$-minimization {\cite{Cappelli:2018vir,Bonanno:2022ztf}}. For a
few intermediate values $d = 3.25, 3.5, 3.75$, islands were found (via scans)
in the three-correlator setup $\langle \sigma \sigma \sigma \sigma \rangle$,
$\langle \sigma \sigma \varepsilon \varepsilon \rangle$, $\langle \varepsilon
\varepsilon \varepsilon \varepsilon \rangle$ {\cite{Behan:2016dtz}}. All these
studies gave results in good agreement with the $\epsilon$-expansion.

Recently, Ref. {\cite{Henriksson:2022gpa}} carried out a systematic study of
the Ising CFT spectrum for $2.6 \leqslant d \leqslant 4$ using the navigator
function $\mathcal{N} (\Delta_{\sigma}, \Delta_{\varepsilon}, \lambda_{\varepsilon \varepsilon \varepsilon} / \lambda_{\sigma \sigma
	\varepsilon})$. Working at $\Lambda = 30$, the low-lying spectrum was
extracted applying the extremal function method at the navigator minimum for
several dimensions in the $2.6 \leqslant d \leqslant 4$ range.

\begin{figure}[h]
\centering
{\includegraphics[width=0.49\textwidth]{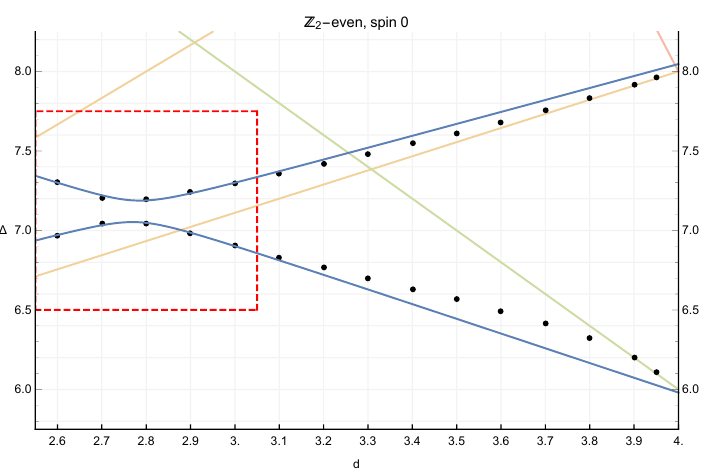}}
	\caption{\label{fig:repulsion}(Color online) Observation of the level repulsion effect
		{\cite{Henriksson:2022gpa}}, \CCBY.  Black dots: spectrum from the bootstrap (extremal functional method). Blue (dark gray) curves: fitted hyperbola to the black dots within the red dashed box. Green (light gray) and orange (light gray) lines: predictions for operator dimensions from the lowest order $4-\epsilon$ expansion.}
\end{figure}

This study led to two lessons. Firstly, it provided further evidence that the
spectrum of the Ising CFT varies continuously with $d$. Secondly, their
important finding was the observation of avoided level crossing between two
scalar $\mathbb{Z}_2$-even operators $\varepsilon''$ and $\varepsilon'''$
which start in $d = 4 - \varepsilon$ with $\Delta'' = 6 + O (\varepsilon)$ and
$\Delta''' = 8 - O (\varepsilon)$. As $d$ is lowered, the lowest order
$\varepsilon$-expansion predicts crossing in $d \approx 3.3$. The perturbative
$\varepsilon$-expansion does not take
into account non-perturbative mixing effects between operators with the same
quantum numbers. Such effects are expected to lead to avoided level crossing
{\cite{Korchemsky:2015cyx,Behan:2017mwi}}, although the precise mechanism of
how this happens when deforming in $d$ is not yet clarified. Ref.
{\cite{Henriksson:2022gpa}} provided evidence for this scenario, observing
that $\Delta''$ and $\Delta'''$ do get close to each other and then repel
around $d \approx 2.78$ with minimal difference $(\Delta''' - \Delta'')_{\min}
\approx 0.136$ (see Fig. \ref{fig:repulsion}).

In the future, it would be interesting to repeat the study of
{\cite{Henriksson:2022gpa}} at higher $\Lambda$, to determine $(\Delta''' -
\Delta'')_{\min}$ more precisely and to gain evidence that their avoided level
crossing is not a finite $\Lambda$ artifact. It would also be very interesting
to develop a theoretical understanding of the non-perturbative mixing effects
which control $(\Delta''' - \Delta'')_{\min}$.

\subsubsection{3-state Potts model: toward the upper critical dimension }

The 3-state Potts model has a second-order phase transition in $d = 2$ while
it has a first-order phase transition in $d = 3$ {\cite{Janke:1996qb}}. It
would be interesting to get a bootstrap proof of this fact, ruling out the
existence of $S_3$ symmetric 3D CFTs which could describe such a transition
{\cite{OpenProblems}}. It is natural to expect that the critical and the
tricritical 3-state Potts CFTs merge and annihilate at some $d_c < 3$. It would
also be interesting to determine $d_c$.\footnote{A
	related problem is to keep $d = 3$ fixed and vary $q$. The critical
	and the tricritical $q$-state Potts CFTs should then merge and annihilate at some $q_c
	< 3$. Old Monte Carlo simulations suggest $q_c \approx 2.45$
	{\cite{LeeKosterlitz}}.}

\begin{figure}[h]
\centering
{\includegraphics[width=0.48\textwidth]{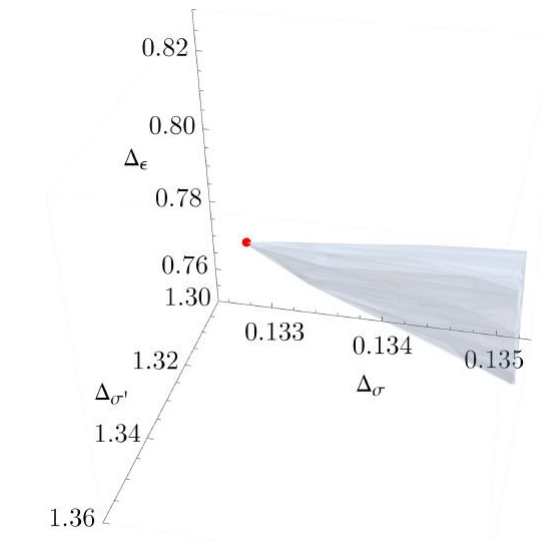}}
	\caption{\label{fig:Potts}(Color online) The cone-shaped allowed region in the space of
		$(\Delta_{\sigma}, \Delta_{\sigma'}, \Delta_{\varepsilon})$ in $d = 2$, $\Lambda =
		11$, whose tip (red dot) closely agrees with the exact solution of the
		critical 3-state Potts model {\cite{Chester:2022hzt}}.}
\end{figure}

Recently, Ref. {\cite{Chester:2022hzt}} attacked the second question using the
navigator function method. Their setup involved all 19 4pt correlation
functions of external operators $\sigma, \sigma'$, $\varepsilon$, where
$\sigma, \sigma'$ are the two relevant scalar primaries in the fundamental
irrep while $\varepsilon$ is the relevant scalar singlet (assumed
the only relevant scalars in these irreps). Their theory space had 10 parameters: 3 scaling dimensions of $\sigma, \sigma'$, $\varepsilon$, and 7 ratios of OPE coefficients among them. Using the Delaunay triangulation/cutting surface algorithm, they found a cone-shaped allowed region in the space of
$(\Delta_{\sigma}, \Delta_{\sigma'}, \Delta_{\varepsilon})$ and identifed the 3-state Potts CFT
with the sharp tip of this region, see Fig.~\ref{fig:Potts}.

They also propose a similar identification for the tricritical 3-state Potts
CFT, using a navigator function setup involving the second relevant singlet
scalar $\varepsilon'$ and a subset of 21 4pt correlation functions involving
$\sigma, \sigma', \varepsilon, \varepsilon'$. In $d = 2$, and at $\Lambda = 11$, both
identification agree very well with the exact solution.

\begin{figure}[h]
	\centering{\includegraphics[width=0.48\textwidth]{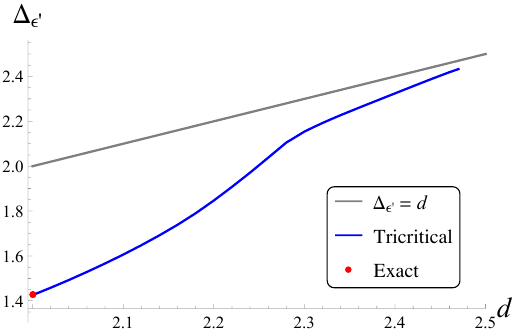}}
	\caption{\label{fig:Potts-merger}(Color online) Tricritical $\Delta_{\varepsilon'}$ as a
		function of $d$ extracted by {\cite{Chester:2022hzt}}.}
\end{figure}

Inspired by this agreement, using the navigator function setups, Ref. {\cite{Chester:2022hzt}} tried to track the
critical and the tricritical CFTs as a function of $d > 2$, assuming the same
identification as in $d = 2$. They wanted to see if the two CFTs come together
and merge. The results of this study were not fully satisfactory. First, while
the tricritical scaling dimensions varied continuously, the critical
dimensions experienced an unexpected discontinuous jump at $d \approx 2.03$. Second,
focusing on the cleaner tricritical curve, one could hope to detect
$d_c$ as the point where $\Delta_{\varepsilon'} (d) = d$. On general
grounds, one expects 
a square-root behavior near this crossing {\cite{Gorbenko:2018ncu}}:
\begin{equation}
	\Delta_{\varepsilon'} (d) - d_c \propto \sqrt{d_c - d} \qquad (d_c - d \ll
	1) .
\end{equation}
Instead, Ref. {\cite{Chester:2022hzt}} observed the behavior in Fig.
\ref{fig:Potts-merger}. We see that $\Delta_{\varepsilon'}$ starts approaching
the $\Delta = d$ line according to the square-root law, but around $d \sim
2.28$ the behavior crosses over to a linear approach.

To overcome these difficulties, it would be desirable to find a bootstrap
setup where the critical and tricritical CFT would be isolated into islands.
Then $d_c$ could be determined from the disappearance of the islands.
Unfortunately, such a setup remains elusive.

\section{Skydive}

In this section we will describe {\tt skydive} \cite{NingSu-skydive, Liu:2023elz}, the latest dramatic improvement in the series of numerical conformal bootstrap technology improvements which started with the introduction of the navigator function. 

\subsection{Basic idea}
\label{sec:skydive-basic}
The typical procedure in navigator computation involves the following steps: (1) select a point in the parameter space and generate the corresponding SDP; (2) compute the SDP to obtain the navigator function and its gradient; (3) make a move in the parameter space based on the local information from the navigator function; and then repeat the process. In the step (2), the SDP needs to be fully solved. In the technical terminology of the SDP algorithm realized in {\tt SDPB} \cite{Simmons-Duffin:2015qma}, an SDP is considered solved when an internal parameter $\mu$, which can be interpreted as an error measure, is reduced below a certain threshold, typically $10^{-30}$. Hot-starting often leads to the stalling of the solver if the checkpoint has a very small $\mu$.\footnote{This is not so surprising since SDPB uses an \emph{interior point} algorithm, which moves through the interior of the allowed region to reach the optimal point on its boundary.} A robust strategy avoids hot-starting.\footnote{One could attempt to save a checkpoint when $\mu$ is not too small. Although the stalling problem is usually less severe in this case, there is still no guarantee that stalling won't occur.} A typical navigator run spends most of its time computing SDPs.

\begin{figure}[!h]
	\centering
	\includegraphics[width=0.55\textwidth]{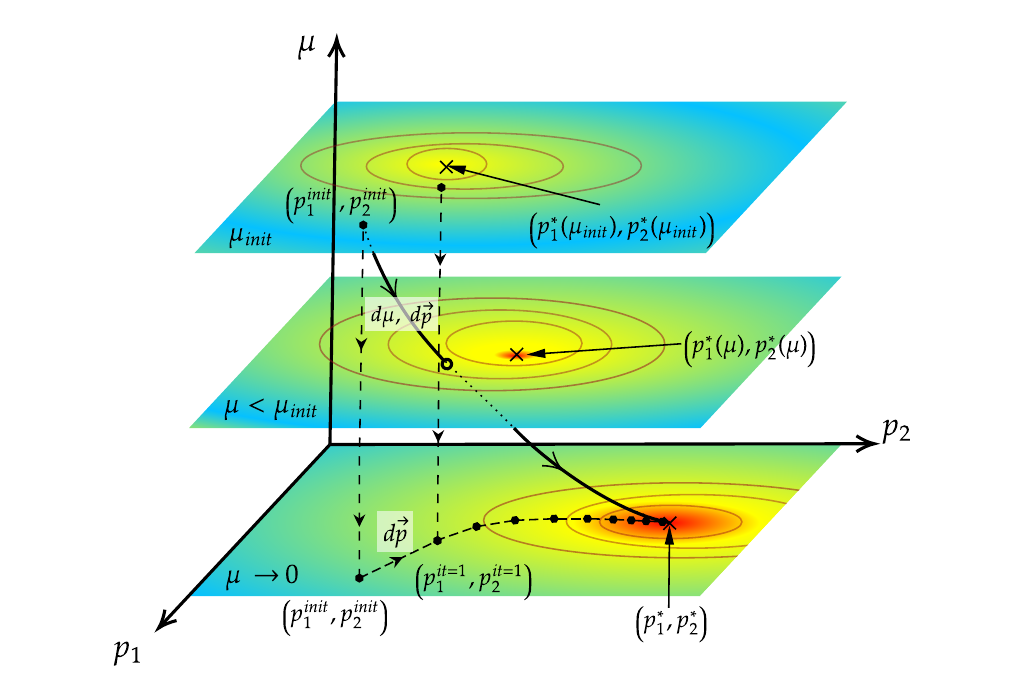}
	\caption{\label{fig:skydive_idea} (Color online) The original navigator function method compared to the skydiving algorithm. The goal is to minimize the navigator function $\mathcal{N}(p)$ where $p=(p_1,p_2)$ represents parameters in theory space. In the original navigator method, we solve an SDP at fixed $(p_1,p_2)$ to get $\mathcal{N}(p)$ at $\mu=0$, following the vertical dashed path. Then, we update $(p_1,p_2)$ (suggested by the BFGS algorithm), compute another SDP starting from $\mu_\text{init}$, and repeat the process. On the other hand, {\tt skydive} changes both $p$ and $\mu$ simultaneously, as depicted by the solid black path, converging to the same navigator minimum. Figure from \cite{Liu:2023elz}.}
\end{figure}

The basic idea of {\tt skydive}  is to optimize this approach, using an intuition that an SDP does not need to be completely solved to obtain a rough estimate of the navigator function value. Indeed, a good estimation of the navigator function can be achieved at a finite $\mu$, and such information is sufficient to indicate a good move in theory space. An ideal scenario is as follows: when the solver is far from the final optimal point, it computes the SDP until the $\mu$ is small enough to provide a reliable estimation of the navigator function, yet not so small as to cause stalling during hot-starting. Based on this estimation, the solver then moves to a new SDP nearby in the parameter space and initiates the computation of this new SDP with the previous checkpoint. In the new computation, only a few iterations are needed to obtain an acceptable estimate of the navigator function, since the checkpoint is essentially almost correct. As the solver progresses toward the final optimal point, we should methodically decrease $\mu$ to refine the estimates of the navigator function, eventually converging on the optimal point with $\mu$ below the threshold. This ideal scenario is illustrated in Figure \ref{fig:skydive_idea}.

\subsection{Algorithm}
\label{sec:skydive-alg}

We will now describe the skydiving algorithm realizing this idea \cite{Liu:2023elz}, which dramatically accelerates navigator-type computations.

The skydiving algorithm can be understood as an upgrade of the primal-dual interior point algorithm used in {\tt SDPB}. We begin with a brief summary of the latter. The SDP can be formulated using the Lagrange function:
\begin{align}
	L_\mu(x,y,X,Y) &= c^T x + b^T y - x^T B y \nonumber \\
&+ \text{Tr}((X - x^T A_*)Y) - \mu \log \det X,
	\label{SDPLagrangian}
\end{align}
where $X, Y \in \mathcal S^K$ (symmetric $K\times K$ matrices), $A_*=(A_1,\ldots,A_P) \in  (\mathcal S^K)^P$, $B \in \mathbb (\mathbb R^{n})^P$ are rectangular $n\times P$ matrices, and $c,x \in \mathbb R^P$, $b,y \in \mathbb R^n$ are vectors. In bootstrap applications, the constant $b,c,B,A$ quantities are related to the bootstrap conditions and conformal blocks, following the procedure in \cite{Simmons-Duffin:2015qma}. In this formalism, the SDP consists in finding the stationary point\footnote{The stationary solution is defined to be a solution to $\partial L/\partial \xi=0$ for $\xi=x,y,X,Y$.} of $L_\mu$ as $\mu \to 0$, subject to $X,Y \succeq 0$. The last term of \eqref{SDPLagrangian} represents a barrier function which pushes us to the interior of the set $X,Y\succeq 0$. This barrier function disappears when the limit $\mu\to0$ is taken. The primal-dual interior point algorithm solves the SDP using a variant of the Newton method with two tweaks: (1) to approach the limit $\mu\to 0$, it gradually decreases $\mu$ by replacing $\mu \to \beta \mu$ in each iteration of the Newton method, with $0 < \beta < 1$; (2) In each step, if the standard Newton steps $dX, dY$ violate the condition $X+dX, Y+dY\succeq 0$, the algorithm performs a partial Newtonian step, rescaling $dX, dY$ with a factor $0<\alpha<1$, to ensure that the positive semidefiniteness of $X, Y$ is preserved.

The user-chosen parameter $\beta$ is important. If $\beta$ is too small, there is a risk of \textit{stalling}, a phenomenon where $\alpha$ decreases to 0, so the steps become shorter and shorter, while the solver is not at the optimal solution. Stalling indicates that the solver is too close to the boundary of the region $X,Y \succeq 0$. On this boundary we have $\det(X)=0, \det(Y)=0$. If $\mu$ is not sufficiently small, movement toward optimality is severely constrained by being close the boundary. As mentioned in Section \ref{sec:skydive-basic}, stalling also frequently occurs when hot-starting a new SDP (i.e.~initializing the solver with given $x, y, X, Y$) using a checkpoint at small $\mu$. This is because one of the stationary conditions is $XY=\mu I$. As $\mu \to 0$, $XY\to 0$, and $X, Y$ become degenerate. From the perspective of the new SDP, the checkpoint is positioned too close to the degeneracy surface while the solution is far from optimal in the new SDP.

Suppose next that we have a family of SDP depending on a parameter $p$ (which may be multi-dimensional). For a fixed $p$ the SDP encodes the computation of a navigator function $\mathcal{N}(p)$, which we want to minimize. This means that we aim to perform optimization not only in $x, y, X, Y$ but also in $p$. We thus extend the Lagrange function to also depend on $p$:
\begin{align}
	\label{pdependentlagrangian}
	L_\mu(x,y,X,Y,p) &= c^T(p) x + b^T(p) y - x^T B(p) y \nonumber \\
&+ \text{Tr}((X - x^T A_*)Y) - \mu \log \det X.
\end{align}
One might have hoped to follow the same idea as in the primal-dual interior point algorithm, i.e.~using the Newton method in the space of $(x,y,X,Y,p)$ and gradually decreasing $\mu$. Unfortunately, this naive approach does not work due to very different roles played by $x, y, X, Y$ and by $p$: (1) In practical conformal bootstrap applications, the dimensions of $x, y, X, Y$ are usually much larger than those of $p$. (2) The Lagrange function depends on $x, y, X, Y$ in a smooth and convex way, but the dependence on $p$ does not have to be convex. In fact it is known that the navigator function can exhibit non-convexity (away from its minimum) and even non-smoothness. If one does try a naive Newton step treating $x, y, X, Y$ and $p$ on equal footing, performance is poor. As the solver moves through a rough landscape in $p$, fixing a constant decreasing rate $\beta$ is a bad idea and often leads to stalling. And when $x, y, X, Y$ are not stationary at a fixed $p$ and $\mu$, the predicted step in $(x,y,X,Y,p)$ is often inaccurate.

Overcoming these difficulties required several new ideas that were introduced in \cite{Liu:2023elz}. (1) The solver dynamically determines a $\beta$ based on the likelihood of stalling and may even temporarily increase $\mu$ (i.e.~use $\beta>1$) when facing a higher risk of stalling. (2) The solver first finds a stationary solution for $x, y, X, Y$ at fixed $p, \mu$, and only then performs a Newton step in $p$. In other words, it attempts to optimize the ``finite-$\mu$ navigator function'', defined as $N(p,\mu)=L_\mu(x^*,y^*,X^*,Y^*,p)$, where $x^*,y^*,X^*,Y^*$ represents the stationary solution for a given $p,\mu$. Since the dimension of $p$ is not too large in practice, the latter optimization is much more manageable than the optimization of $L_\mu(x,y,X,Y,p)$. Moreover, solving the stationary solution for $x, y, X, Y$ at a fixed $p, \mu$ is, in fact, inexpensive. Ref.~\cite{Liu:2023elz} developed a technique that accelerates this part of the computation, typically requiring fewer than four Newton iterations to find a solution with the desired accuracy.

The above discussion would apply generically, when one extremize within a family of SDPs. However, additional difficulties arise when specializing to conformal bootstrap problems. A notable feature of the navigator function in typical conformal bootstrap studies is that it's fairly flatness within the bootstrap island. The gradient inside and outside the island can differ by many orders of magnitude. This feature already poses a challenge in the applications of the navigator function using the original method of \cite{Reehorst:2021ykw}.\footnote{For example, in \cite{Su:2022xnj}, there was a situation where the ordinary BFGS algorithm couldn't effectively handle the sudden change in the order of magnitude of the gradient, and it had to be modified to be efficient.} For the skydiving algorithm, this feature would pose a serious challenge to finding the boundary of an island, because the Lagrange function won't detect the island (defined by $L<0$) until $\mu$ is small enough, contrary to the general idea of skydiving algorithm to move toward the optimal point at relatively big $\mu$. To overcome this challenge, \cite{Liu:2023elz} proposed a modification of the Lagrangian \eqref{pdependentlagrangian} at $\mu>0$, which leads to the same allowed island and the same navigator minimum as $\mu\to0$, but speeds up the optimization algorithm. This modification worked well in the tested examples. See \cite{Liu:2023elz} for the details.

Ref.~\cite{Liu:2023elz} tested the skydiving algorithm on two previously studied problems involving multiple four-point functions: the 3D Ising correlators of $\sigma, \epsilon$, and the O(3) model correlators of lowest vector, singlet, and rank-2 tensor primaries. The results showed that the skydiving algorithm improved computational efficiency in these examples, compared to use of the navigator function without hot-starting, by factors of 10 to 100.

The skydiving algorithm has already been used in several bootstrap studies. Notably, Ref.~\cite{Rong:2023owx} investigated the O(3)-symmetric correlators of $v$ (vector), $s$ (singlet), $t_2$ and $t_4$ scalar primaries (where $t_k$ is a traceless symmetric $k$-index irrep). The information extracted from this computation was used to perform conformal perturbation theory around the O(3) fixed point, to predict the critical exponents of the cubic fixed point. This work also numerically confirmed the large charge expansion prediction $\Delta_{Q,l}=c_{3/2} Q^{3/2}+c_{1/2} Q^{1/2} -0.094 +\sqrt{\frac{l(l+1)}{2}}$ \cite{Hellerman:2015nra, Monin:2016jmo}, including for the first time the spin dependent term comparing $l=0,2$. The setup in \cite{Rong:2023owx} involved the largest number of crossing equations ever explored using the numerical bootstrap: 82 equations. The computations were completed in about 10 days with the skydiving algorithm. In comparison, it would have taken more than a year using the original navigator function method of \cite{Reehorst:2021ykw} (without hot-starting). 

Another application of skydive was Ref.~\cite{Chester:2023njo}, which considered SO(5)-symmetric four-point functions of $v$, $s$, and $t_2$ to investigate the scenario where DQCP is governed by a tricritical point corresponding to a unitary CFT, see Section \ref{sec:bQED}.

Finally, the recent work \cite{ono2} used {\tt skydive} to study multiscalar CFTs with $O(N)\times O(2)$ global symmetry. In this work {\tt skydive} is used to find the minimal value of the parameter $N$ beyond which the unitary $O(N)\times O(2)$ symmetric fixed point ceases to exist.

Clearly, the skydiving algorithm has opened new opportunities in the numerical conformal bootstrap. 
The algorithm of \cite{Liu:2023elz} represents the first attempt at efffective optimization in the $x, y, X, Y, p$ space. Although the algorithm worked for the tested examples, its robustness needs further study,\footnote{The navigator functions for different bootstrap setups may have different features. Some might be more singular than others. Some might have multiple local minima. The difficulty of the optimization also depends on the initial point. We therefore feel that the skydiving algorithm needs more testing in various scenarios.} and future improvements are welcome.

\section{Omissions}

Unfortunately, in our review we could not describe or mention every single numerical conformal bootstrap result obtained since the previous review \cite{Poland:2018epd}. Notable omissions include: 
work on boundary CFTs \cite{Behan:2020nsf,Behan:2021tcn}; solutions of the truncated bootstrap equations (Gliozzi's method \cite{Gliozzi:2013ysa}) in situations where there is no positivity, for 4-point functions \cite{Rong:2020gbi,Nakayama:2021zcr} and for 5-point functions \cite{Poland:2023vpn}; work on superconformal field theories \cite{Bissi:2020jve,Gimenez-Grau:2019hez,Binder:2020ckj,Agmon:2019imm,Lin:2019vgi,Gimenez-Grau:2020jrx,Bissi:2021rei,Liendo:2018ukf,Chester:2021aun,Chester:2022sqb};
conformal bootstrap in situations where the spectrum is known but OPE coefficients need to be determined: \cite{Cavaglia:2021bnz,Cavaglia:2022qpg,Niarchos:2023lot} and \cite{Picco:2019dkm,Nivesvivat:2020gdj,Grans-Samuelsson:2021uor,Ribault:2022qwq}; bounds on CFT correlators \cite{Paulos:2021jxx}; work on other numerical methods such as the extremal flow method \cite{Afkhami-Jeddi:2021iuw}, outer approximation and the analytic functional basis \cite{outerlimit, Ghosh:2023onl}, Machine Learning bootstrap \cite{Kantor:2021jpz,Lal:2023dkj}, stochastic optimization \cite{LuvianoValenzuela:2022yli,Laio:2022ayq}. Regrettably, we also could not describe the work in closely related fields: the modular bootstrap for 2D CFTs  \cite{Lin:2019kpn,Lin:2021udi,Lin:2023uvm,Grigoletto:2021zyv,Grigoletto:2021oho,Chiang:2023qgo,Fitzpatrick:2023lvh,Huang:2019xzm,Bae:2018qym,Bae:2018yre,Hartman:2019pcd,Benjamin:2020zbs,Benjamin:2022pnx,Collier:2021ngi,Afkhami-Jeddi:2020hde,Lanzetta:2022lze};
	bootstrap for Laplacian eigenvalues on surfaces \cite{Bonifacio:2020xoc,Bonifacio:2021msa,Kravchuk:2021akc,Bonifacio:2021aqf,Bonifacio:2023ban,Gesteau:2023brw} (as first shown in \cite{Kravchuk:2021akc}, for hyperbolic surfaces this is extremely closely related to the conformal bootstrap). We could not describe significant progress being achieved in the sister field of the numerical S-matrix bootstrap \cite{Kruczenski:2022lot}. There appeared also several other ``bootstraps" where, like in the  conformal bootstrap, positivity is used to get bounds on spaces of solutions (although the analogy with the conformal bootstrap is less complete due to the absence of analogues of families of conformal blocks). Those deserve separate reviews, and we list but a few entry points into the literature: the lattice model bootstrap \cite{Anderson:2016rcw,Kazakov:2022xuh,Cho:2022lcj}, the matrix model bootstrap \cite{Lin:2020mme,Kazakov:2021lel}, the quantum mechanics bootstrap \cite{Han:2020bkb,Lin:2023owt}.

\section{Discussion and outlook}

In this review, we have covered some advances in numerical conformal bootstrap techniques over the last few years. The development of highly efficient software for computing conformal blocks, effectively solving SDPs, and search algorithms in theory space has been very fruitful. With this progress in numerics, much interesting work has been done. In particular, 3D super-Ising, O(2), O(3), and GNY models have been solved to obtain high-precision critical exponents with rigorous error bars.

Many questions remain to be investigated in the future. There are many CFT targets that we would like to solve. How can we bootstrap critical gauge theories to obtain precise CFT data? Which correlators are the most constraining? How to numerically explore correlators that involve Wilson lines? How can we bootstrap multi-scalar CFTs beyond the Ising and Ising-like theories? What are the most effective gap assumptions? Can we bootstrap complex CFTs? These are still wide open questions.

Interestingly, these questions are open even for the 2D minimal models. Can we find a systematic way to constrain minimal models into small bootstrap islands? If so, we might hope to extend these islands into non-integer dimensions $d>2$.

To bootstrap these and other targets one may think of, it is likely that we will need to further develop the bootstrap machinery to make it ever more powerful. Given a set of correlators, can we make the bootstrap bound converge faster? This question is especially crucial for correlators of large dimension operators (as is the case for critical gauge theories) since they often converge slowly in the current computational framework. There are several possibilities for acceleration:
\begin{itemize} 
	\item 
	We may hope to accelerate the higher $\Lambda$ computation with information from a lower $\Lambda$ computation. This has been employed in the 2D chiral modular bootstrap and has achieved amazing acceleration in computation \cite{Afkhami-Jeddi:2019zci}. 
	\item
    In various limits, analytic understandings of the spectrum have been achieved, such as the lightcone limit and large charge limit for O(N). Combining this analytic information with numerical bootstrap should further strengthen the constraining power. A non-rigorous ``hybrid bootstrap'' method to combine the numerical and the lightcone information has been explored in \cite{Su:2022xnj}. It would be interesting to explore more robust and rigorous methods. 
    
    \item Instead of using the derivative basis, there are better bases that could converge faster. By using a better basis, amazing improvement has been achieved in the 1D CFT \cite{Paulos:2019fkw,Ghosh:2021ruh} and 2D CFT case \cite{Ghosh:2023onl}. In particular in the latter work the authors observed that the bound at large external dimensions converges significantly better than what can be practically achieved with the derivative basis. Overcoming the remaining challenges to extend this method to 3D would be very useful for bootstrapping the targets mentioned in the previous paragraphs.
    
    \end{itemize}

We remain optimistic that numerical bootstrap technology will continue seeing significant improvements in the future, and we hope that these advancements will continue yielding fascinating insights in physics questions.

\section*{Acknowledgements}
We thank the numerical conformal bootstrap community for the effort which led to the development of the ideas and results presented in this review. We are particularly grateful to our collaborators Shai Chester, Yin-Chen He, Aike Liu, Junyu Liu, Walter Landry, David Poland, Junchen Rong, Marten Reehorst, Benoit Sirois, David Simmons-Duffin, Balt van Rees, Alessandro Vichi. We thank David Simmons-Duffin for communications about the {\tt tiptop} algorithm, and Johan Henriksson and Stefanos Kousvos for discussions and comments. SR is supported by the Simons Foundation grant 733758 (Simons Collaboration on the Nonperturbative Bootstrap). Most of NS's work was conducted at the University of Pisa, where this project has received funding from the European Research Council (ERC) under the European Union's Horizon 2020 research and innovation programme (grant agreement no. 758903). NS's work at the California Institute of Technology is supported in part by Simons Foundation grant 488657 (Simons Collaboration on the Nonperturbative Bootstrap).

\appendix

\section{Comment on islands, isolated regions, and kinks}
\label{app:comment}
In this review, we use \textit{island}, \textit{isolated/closed regions}, and \textit{kinks} to describe various bootstrap results. We would like to clarify how we use those terms.

We call an isolated allowed region in the CFT parameter space a \textit{bootstrap island of theory X} if it satisfies the following conditions: (1) The gap assumptions are mild (i.e.~not too close to the actual value in theory X) and/or naturally motivated (for example, by the number of relevant operators, or by equations of motion creating gaps in the spectrum, etc); (2) The isolated region is insensitive to the gap assumptions, i.e., the region doesn't change much if the gaps are varied slightly; (3) The isolated region is small enough that it does not contain any known theories other than X. For example, we consider the 3D Ising CFT island in Figure 1 of \cite{Kos:2016ysd} to satisfy these conditions. When a bootstrap island is obtained, we often can determine some CFT data with more or less rigorous error bars, by scanning over the island. One of the main goals of the conformal bootstrap program is to find the correct setups to isolate various CFTs into islands.

\emph{Kinks} are sudden changes of the slope of the boundary of a feasible region. We call a kink a \textit{stable kink of theory X} to mean that (1) The gap assumptions leading to the kink are mild and naturally motivated, and (2) the location of the kink is insensitive to the gap assumptions and is consistent with what is known or conjectured about the position of theory X. For example, the kink in Figure 5 of \cite{El-Showk:2012cjh} is a stable 3D Ising CFT kink because, when the gap changes slightly, the location remains the same, although the sharpness of the kink may change. With a stable kink, we often could make prediction on some CFT data, although the error bars may not be as rigorous as with the islands. Sometimes, one observes \textit{moving kinks}, whose location changes with the gap assumptions (such as the example discussed in Section \ref{sec:bQED}). Those are obviously less useful than the stable kinks, but not totally useless. Indeed, in some examples, it is expected that when a certain gap is set to the actual CFT value, the CFT would precisely saturate the kink's moving position. In these cases, bootstrap computation could make predictions on CFT data once a few pieces of information obtained using other methods are injected.

One should be careful when considering large-$N$ CFT islands. Different CFTs could have the same CFT data in the $N\to \infty$ limit. Therefore, in the theory space, there could be multiple solutions that are very close to each other at large-$N$, which may not be easily distinguishable by numerical bootstrap. These solutions may show a significant split when $N$ becomes small. Consequently, apparent success at large-$N$ does not automatically translate to success at smaller $N$. For example, in the case of the bosonic QED$_3$ feasible region shown in FIG. 3 of \cite{He:2021xvg}, the region is closed for sufficiently large-$N$, but not at small $N$. It's unclear whether this is purely due to a convergence issue\footnote{It's known that convergence in $\Lambda$ is slow for correlators of operators with large scaling dimensions, which is the case here.} or if there are several CFTs satisfying the imposed conditions.

Even though the ultimate goal of the conformal bootstrap program is to solve CFTs, for many target theories, it's not yet clear how to obtain islands or stable kinks. In those cases, people compute generic bounds, while the target CFTs are not on the boundary of the bounds. These bootstrap bound are also very valuable, since sufficiently strong bounds could offer insight into related physics questions (for example, as in the case of DQCP, reviewed in Section E of \cite{Poland:2018epd}).

\bibliography{nbr-biblio}
\bibliographystyle{utphys}

\end{document}